\documentclass[lettersize,journal]{IEEEtran}
\usepackage{amsmath,amsfonts}
\usepackage{algorithmic}
\usepackage{algorithm}
\usepackage{array}
\usepackage[caption=false,font=normalsize,labelfont=sf,textfont=sf]{subfig}
\usepackage{textcomp}
\usepackage{stfloats}
\usepackage{url}
\usepackage{verbatim}
\usepackage{graphicx}
\usepackage{cite}
\usepackage[hidelinks]{hyperref}
\usepackage{bm}
\usepackage{tabularx}
\usepackage{multirow}
\usepackage{xcolor}
\usepackage{float}
\usepackage{units}

\begin{document}
\begin{sloppypar}

\setlength{\textfloatsep}{5pt}  
\setlength{\floatsep}{5pt}      
\setlength{\intextsep}{5pt}     
\setlength{\belowcaptionskip}{3pt} 
\setlength{\abovecaptionskip}{5pt} 

\title{Non-Iterative Coordination of Interconnected Power Grids via Dimension-Decomposition-Based Flexibility Aggregation}

\author{Siyuan Wang,~\IEEEmembership{Member,~IEEE,}
        Cheng Feng,~\IEEEmembership{Member,~IEEE,}
        Fengqi You,~\IEEEmembership{Senior Member,~IEEE}
\thanks{Manuscript received xx, xx, 2025. Paper no. TPWRS-00234-2025. (\textit{Corresponding author: Fengqi You.})}
\thanks{S. Wang, C. Feng and F. You are with College of Engineering, Cornell University, Ithaca, New York 14853, USA. (Email: \{siyuan.wang, chengfeng, fengqi.you\}@cornell.edu).}}

\markboth{IEEE Transactions on Power Systems,~Vol.~XX, No.~XX, XX~202X}%
{Wang \MakeLowercase{\textit{et al.}}: Non-Iterative Coordination of Interconnected Power Grids via Dimension-Decomposition-Based Flexibility Aggregation}


\maketitle

\begin{abstract}

The bulk power grid is divided into regional grids interconnected with multiple tie-lines for efficient operation. 
Since interconnected power grids are operated by different control centers, coordinating dispatch across multiple regional grids is challenging. 
A viable solution is to compute a flexibility aggregation model for each regional power grid; then optimize the tie-line schedule using the aggregated models to implement non-iterative coordination. 
However, challenges such as intricate interdependencies and the curse of dimensionality persist in computing the aggregated models in high-dimensional space. 
Existing methods like Fourier-Motzkin elimination, vertex search, and multi-parameter programming are limited by dimensionality and conservatism, hindering their practical application. 
This paper presents a novel dimension-decomposition-based flexibility aggregation algorithm for calculating the aggregated models of multiple regional power grids, enabling non-iterative coordination in large-scale interconnected systems. 
Compared to existing methods, the proposed approach yields a significantly lower computational complexity and a less conservative flexibility region. 
The derived flexibility aggregation model for each regional power grid has a well-defined physical counterpart, which facilitates intuitive analysis of multi-port regional power grids and provides valuable insights into their internal resource endowments.
Numerical tests validate the feasibility of the aggregated model and demonstrate its precision in coordinating interconnected power grids.

\end{abstract}

\begin{IEEEkeywords}
Regional power grid, flexibility aggregation, interconnected systems, dimension decomposition, polytope projection.
\end{IEEEkeywords}


\section*{Nomenclature}

\subsection*{Variables}

\begin{tabularx}{0.47\textwidth}{lX}
	$\bm{x}^\text{bd}$, $\bm{x}^\text{int}$  & Vectors composed of all the boundary variables and internal variables of all time slots.\\
	$\bm{p}^\text{tpg}_t$, $\bm{p}^\text{rgu}_t$  & Vectors composed of thermal power generators and renewable generators' output power at time $t$.\\
	$Rd^\text{tpg}_{i,t}, Ru^\text{tpg}_{i,t}$ & Down and up reserve of $i$-th thermal power generator at time $t$.\\ 
	$\bm{p}^\text{tie}_t$ & Vector composed of tie-lines' output power at time $t$. \\
	$\bm{p}^\text{ld}_t$ & Vector composed of buses' load power at time $t$.
\end{tabularx}

\begin{tabularx}{0.47\textwidth}{lX}
	$\bm{\theta}_t$ & Vector composed of the voltage angles of all buses at time $t$. \\
	$\bm{\theta}^\text{bd}_t$ & Vector composed of the voltage angles of boundary buses that connected to the AC tie-lines at time $t$.
\end{tabularx}

\subsection*{Notations and Operators}

\begin{tabularx}{0.48\textwidth}{lX}
	$\mathcal{R}_{\text{H}}$ & The feasible region representing all the operational constraints of a system within a high-dimensional space where all decision variables reside.\\
	$\mathcal{R}_\text{L}$ & The aggregated flexibility region of a system within the projected space where the boundary variables reside. \\
	$\hat{\mathcal{R}}_\text{L}$ & The inner-approximated aggregate flexibility region of the system, characterized by a specific shape template. \\
	$(\cdot)_i$ & The $i$-th element of a vector or the $i$-th row vector of a matrix. \\
	$\overline{(\cdot)}, \underline{(\cdot)}$ & The upper- and lower-bound parameters of variables. \\
	$[n]$ & The set of integers from 1 to $n$.  \\
	$\underset{i \in \mathcal{N}}{\operatorname{col}}\left(x_i\right)$ & The column vector composed of elements $x_i$ for all indices $i \in \mathcal{N}$. \\
	$\operatorname{space}(\bm{x})$ & The space where the vector $\bm{x}$ resides. $\operatorname{space}(\bm{x})=\mathbb{R}^{n}$, where $n:=\operatorname{len}(\bm{x})$ \\
	$\operatorname{card}(\mathcal{T})$ & The cardinality of the set $\mathcal{T}$.
\end{tabularx}


\section{Introduction}

\subsection{Background and Motivation}

\IEEEPARstart{T}{he modern} power grids in China, North America and Europe are all interconnected systems. To facilitate administration and coordination, the bulk power grid is segmented into several regional power grids (RPGs) based on geographical areas and interconnected by AC and DC tie-lines \cite{zhangCoordinatedSchedulingGenerators2021}. 
The proportion of renewable energy and the varying weather conditions in various regions create complementary patterns in peak electricity supply and demand.
Energy transmission through AC and DC tie-lines provides cross-regional energy support and ensures stable and economical operation of the whole system.

However, achieving optimal coordinated dispatch across multiple RPGs presents significant challenges, as each RPG operates under its own control center. 
Data privacy concerns and variations in data models further complicate the coordination between these grids. 
In addition, the sheer volume of devices and network data makes centralized management and optimization of the entire bulk power grid impractical. 
A commonly adopted approach is to coordinate RPGs using a decentralized optimization framework; the optimal dispatch problem of each RPG is then treated as a sub-problem, with the overall coordination of the bulk power grid achieved through an iterative process.
An iteration-based decomposition method, with quasi-second-order convergence rate, proposed in \cite{linDecentralizedDynamicEconomic2018}, generalizes from linear to non-linear problems. This represents a substantial advancement in decomposition techniques, significantly reducing the number of iterations and computational burden at each control center.
Nevertheless, guaranteeing convergence in such iterative schemes remains challenging, and diagnosing and remedying divergence when it arises can be difficult. 
Additionally, coordinated dispatch requires synchronized real-time communication between all regional control centers, which creates significant obstacles in practical engineering implementation.

\begin{figure}[h] 
	\centering 
	\includegraphics[width=0.48\textwidth]{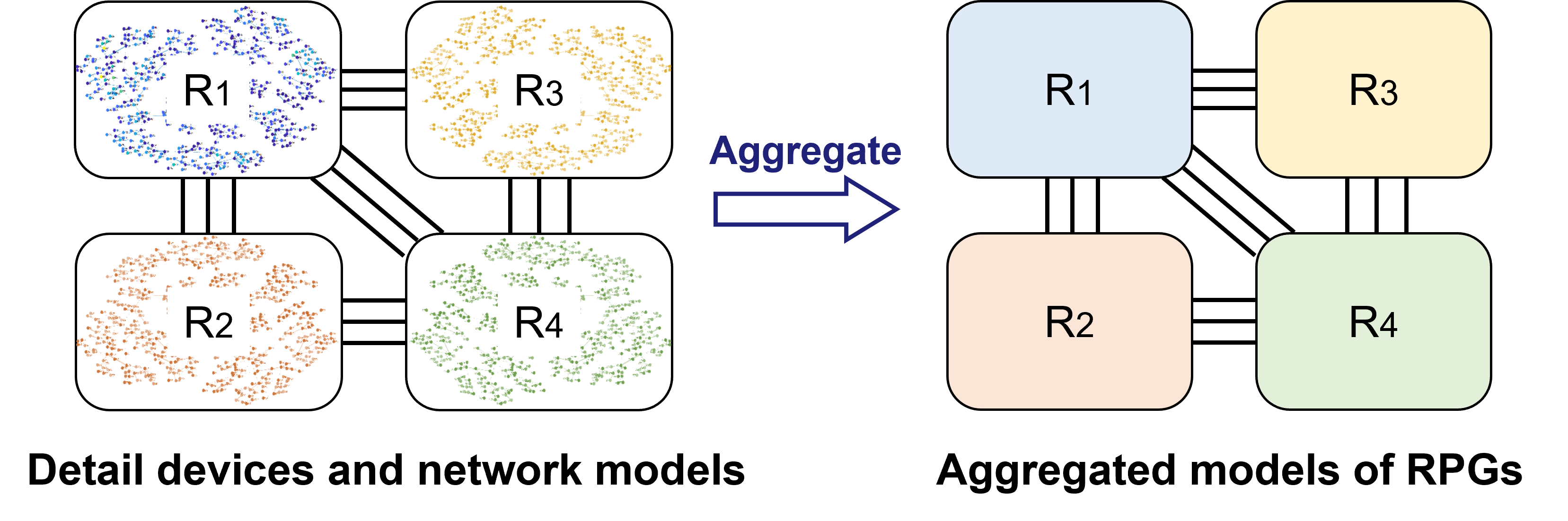}
	\caption{Aggregation of regional power grids for coordinated dispatch of interconnected systems.} 
	\label{fig-aggMultiRegionalGrid}
\end{figure}

An effective solution to these challenges is to calculate the power transmission flexibility region of tie-lines between RPGs, which is also called tie-line security region in \cite{linTieLineSecurityRegion2021,linHighdimensionTielineSecurity2023}.
Then coordination can be achieved via a non-iterative approach \cite{tanNonIterativeSolutionCoordinated2024}. 
A schematic diagram illustrating the aggregation of regional power grids is shown in Fig.~\ref{fig-aggMultiRegionalGrid}. 

By employing an aggregated model, each RPG can focus solely on the external characteristics of tie-lines, without considering complex internal network constraints, thereby significantly reducing the complexity of the optimal coordination of the bulk grid.
The results of flexibility aggregation can also be used as a comprehensive assessment of the overall RPG, providing valuable information for the endowment of resources within the system \cite{wangAggregationReferenceModel2024}.
At the same time, device data and information within each RPG is effectively protected through aggregation.

A comprehensive comparison between distributed iterative and aggregated model-based non-iterative methods is summarized in TABLE~\ref{tab-comparison}.

\begin{table}[h]
    \centering
    \footnotesize
    \caption{Comparison Between Distributed Iterative and Aggregated Model-Based Non-Iterative Methods}
    \renewcommand{\arraystretch}{1.5}
    \begin{tabularx}{0.45\textwidth}{>{\centering\arraybackslash}m{0.09\textwidth}>{\centering\arraybackslash}m{0.18\textwidth}>{\centering\arraybackslash}m{0.12\textwidth}}
        \hline\hline
        Characteristic & Iterative & Non-Iterative \\
        \hline
        Convergence & Not guaranteed; failure causes are often unclear & Guaranteed in a single step \\
        Iteration count & Multiple (tens to hundreds) & One-time computation \\
        Communication & Requires real-time, synchronous updates per iteration; vulnerable to bottlenecks from the slowest participants & One-time aggregated model exchange \\
        Real-time coordination computational burden & High cumulative runtime & Moderate one-time overhead \\
        Scalability & Iterations increase with the number of sub-problems & Highly scalable due to decoupling \\
        Implementation & Requires complex coordination protocols & Simple and modular deployment \\
        \hline\hline
    \end{tabularx}
    \label{tab-comparison}
\end{table}

\subsection{Literature Review}

Calculating an RPG’s aggregated model still presents major challenges. 
Firstly, the decision variables in an RPG are numerous. 
All these variables are coupled with each other by large-scale spatio-temporal coupling constraints, including the network constraints at single time slices and the temporal coupling constraints of generation units across all time slots. 
All the variables and the coupling constraints form a very complex, high-dimensional space.
In addition, an RPG connects to other RPGs through multiple AC and DC tie-lines, which network constraints further interconnect. 
It is also difficult to explicitly express these spatio-temporal coupling relationships as constraints among these tie-lines.

Some existing studies have proposed several types of methodologies for computing the flexibility aggregation model of a system. 
The Fourier-Motzkin elimination method is proposed in \cite{dantzigFourierMotzkinEliminationIts1973,a.a.jahromiLoadabilitySetsPower2017} to calculate the flexibility region of the power system by eliminating the internal variables of the system.
The concept of the dispatchable region is presented in \cite{weiDispatchableRegionVariable2015} and calculated by iteratively creating boundary hyperplanes of the feasible region. 
The progressive vertex enumeration method is proposed in \cite{tanEnforcingIntraRegionalConstraints2019} to calculate the projected flexibility region of the power system by iteratively searching for the extreme points. 
In addition, the work in \cite{linTieLineSecurityRegion2021} presents an evaluation method for the secure region of a single tie-line between regional power grids. 
The multi-parameter programming-based algorithms are presented in \cite{linDeterminationTransferCapacity2019}, and an enhanced acceleration scheme is proposed in \cite{linTieLinePowerTransmission2020}. 
The security region of renewable energy integration is calculated in \cite{daiSecurityRegionRenewable2019} with multi-parameter programming.
Geometry-based methods also offer an approach to calculate the aggregation model, where the aggregated flexibility region is approximated using various geometric shapes, such as cubes \cite{chenLeveragingTwoStageAdaptive2021}, ellipsoids \cite{cuiNetworkCognizantTimeCoupledAggregate2021}, and polytopes shaped by storage and/or generator constraints \cite{zhaoAggregatingAdditionalFlexibility2020,wangAggregateFlexibilityVirtual2021,wenAggregateFeasibleRegion2022}.
Moreover, to deal with the high-dimensional coupling among multiple tie-lines in an RPG, \cite{linHighdimensionTielineSecurity2023} decomposes the flexibility region along the temporal dimension and recombines the lower-dimensional regions with the Cartesian product. 

In practical power grids, the presence of multiple tie-lines between regional grids creates a multi-port system, significantly increasing computational complexity. 
Existing methods such as Fourier-Motzkin elimination, vertex search, multi-parameter programming, and direct geometric projection all suffer from the curse of dimensionality. 
As a result, these approaches often fail to compute the flexibility region of RPGs in a limited time, limiting their practical application in engineering.  
Furthermore, existing methods exhibit excessive conservatism, producing flexibility regions that are much smaller than the actual regions.
This limitation prevents them from achieving optimal solutions in dispatch applications and hinders their ability to attain economic optimality.

\subsection{Contributions and Paper Organization}

This paper proposed a novel dimension-decomposition-based inner approximation method to calculate the flexibility aggregation model of a regional power grid with multiple AC and DC tie-lines. 
The aggregated model integrates the network constraints, devices constraints, $N$-1 security constraints, and the capability of generators within each regional power grid. 
With the flexibility aggregation model of regional power grids, it can greatly simplify and accelerate the schedule of tie-lines and coordination of interconnected regional power grids in a non-iterative way.

The contributions of this paper are outlined as follows:

(1) A flexibility aggregation model is developed for each regional power grid in the bulk power system, incorporating network constraints, device constraints, and $N$-1 security constraints. These aggregated models significantly simplify and accelerate the scheduling of tie-lines, and facilitate the coordination of interconnected power grids using a non-iterative approach.

(2) A novel dimension-decomposition-based flexibility aggregation algorithm is proposed to calculate the flexibility region of each regional power grid. This method not only accelerates the computation process, making the high-dimensional problem tractable, but also achieves substantially less conservative flexibility aggregation results compared to existing approaches.

(3) The derived flexibility aggregation model of each regional power grid has a well-defined physical counterpart, which facilitates intuitive analysis of multi-port regional power grids and provides valuable insights into their internal resource endowments.

The remainder of this paper is organized as follows. 
Section \ref{sec-Methodology} provides an interpretation and methodology for the flexibility aggregation problem. 
Section \ref{sec-RPGMathModel} introduces the mathematical model for a regional power grid. 
In Section \ref{sec-RPGAgg}, the dimension-decomposition-based inner approximation algorithm is presented to calculate the flexibility aggregation model. Physical interpretation of the flexibility aggregation model is also given.
Section \ref{sec-RPGAggCoordDispatch} describes the coordinated dispatch model of interconnected RPGs based on the aggregated flexibility results. 
Numerical tests are conducted in Section \ref{sec-NumericalTest}, followed by conclusions in Section \ref{sec-Conclusion}.


\section{Preliminaries: Interpretation and Methodology for the Flexibility Aggregation}
\label{sec-Methodology}

\subsection{Physical and Mathematical Interpretation for the Flexibility Aggregation Problem}

\begin{figure}[h] 
	\centering 
	\includegraphics[width=0.48\textwidth]{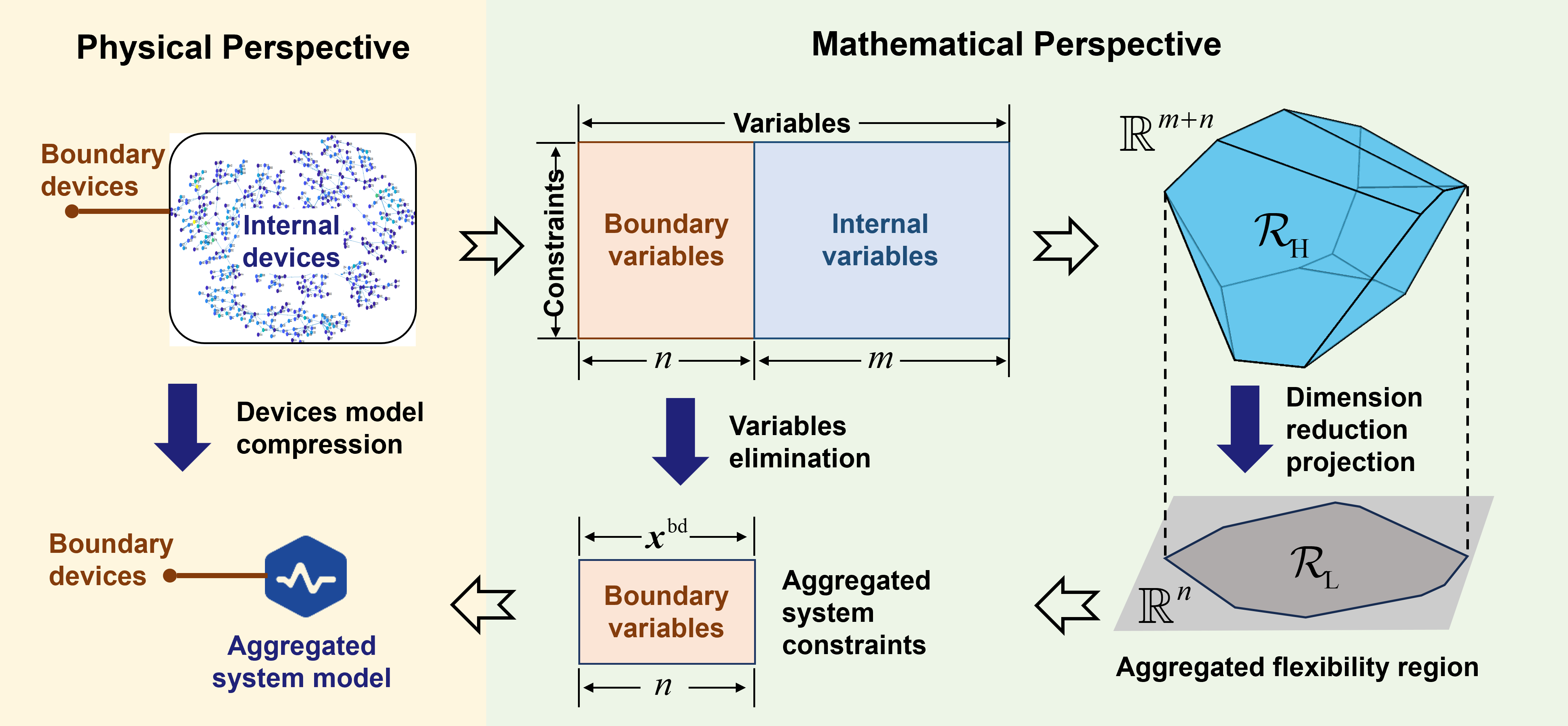}
	\caption{Physical and mathematical perspectives of system flexibility aggregation.} 
	\label{fig-ThreePerspective}
\end{figure}

System flexibility is defined as the capacity to maintain a secure and cost-effective supply-demand balance across spatial and temporal scales, achieved through the seamless coordination of various controllable assets \cite{dallaneseUnlockingFlexibilityIntegrated2017,chiccoFlexibilityDistributedMultienergy2020}. 
The flexibility aggregation of a system can be viewed from both the mathematical and physical perspectives, as shown in Fig.~\ref{fig-ThreePerspective}.

From a physical perspective, a system consists of numerous interconnected internal devices. These devices influence each other and interact with the external environment through boundary devices, such as feeders or tie-lines, enabling energy and information exchange.
To aggregate the flexibility of the entire energy system, the internal devices and coupling networks are compressed into a flexible equivalent aggregation model.
Then, the system's overall flexibility, viewed externally, can be represented solely by the boundary devices. The internal structure of this system is treated as a black box.

Considering the mathematical model of all the devices in the system, the system is governed by the operational and security constraints involving the internal variables $\bm{x}^\text{int} \in \mathbb{R}^{m}$ and the boundary variables $\bm{x}^\text{bd} \in \mathbb{R}^{n}$. 
If these constraints can be approximated or relaxed in a linear form, they constitute the feasible region $\mathcal{R}_{\text{H}}$ of the whole system as follows:

\begin{equation} \label{eq-original_polytope}
	\mathcal{R}_{\text{H}}=
	\left\{
		\left[
			\begin{array}{l}
				\bm{x}^\text{bd} \\
				\bm{x}^\text{int}
			\end{array} 
		\right]
		\left\lvert\,
			\begin{array}{l}
				\bm{x}^\text{bd} = \bm{C} \bm{x}^\text{int} + \bm{d}  \\
				\bm{F} \bm{x}^\text{int} \leq \bm{f}
			\end{array}  
		\right.
	\right\}
\end{equation} 

$\mathcal{R}_{\text{H}}$ can also be viewed as a polytopal region in the high-dimensional space $\mathbb{R}^{m+n}$. 
To aggregate the flexibility of the system, all internal variables $\bm{x}^\text{int}$ should be eliminated from the constraints of the system, so that the aggregated flexibility region can be expressed only by the constraints of the boundary variables $\bm{x}^\text{bd}$. 
From a geometric perspective, this is equivalent to the projection of the polytope $\mathcal{R}_{\text{H}}$ onto the lower-dimensional space $\mathbb{R}^{n}$, where the boundary variables $\bm{x}^\text{bd}$ reside:

\begin{equation} \label{eq-projection_operation}
	\mathcal{R}_\text{L}=
	\left\{ 
		\bm{x}^\text{bd} 
		\left\lvert\
		\begin{array}{l}
			\forall \bm{x}^\text{bd} \in \mathcal{R}_\text{L}, 
			\exists \bm{x}^\text{int}: \\
			\left[
				\begin{array}{l}
					\bm{x}^\text{bd} \\
					\bm{x}^\text{int}
				\end{array} 
			\right] \in \mathcal{R}_{\text{H}}
		\end{array}
		\right.
	\right\}
\end{equation}

Since the dimensionality reduction projection of a polytope can retain the form of a polytope, the projected polytope can be expressed as

\begin{equation} \label{eq-projected_polytope}
	\mathcal{R}_\text{L}=
	\left\{
		\bm{x}^\text{bd} 
		\left\lvert\
			\bm{M}^{*} \bm{x}^\text{bd} \leq \bm{n}^{*} 
		\right. 
	\right\}, 
	\mathcal{R}_\text{L} \subset \mathbb{R}^{n} 
\end{equation}

\noindent where matrix $\bm{M}^{*}$ and vector $\bm{n}^{*}$ are obtained by projection operation.
However, the system's large number of decision variables creates a high-dimensional space, leading to an over-exponential increase in computational complexity and rendering the problem NP-hard \cite{zhenComputingMaximumVolume2018,wenAggregateFeasibleRegion2022}.
Consequently, precise calculation methods to obtain exact values of $\bm{M}^{*}$ and $\bm{n}^{*}$, such as Fourier-Motzkin elimination \cite{dantzigFourierMotzkinEliminationIts1973} and polytope projection algorithms, become intractable. 

\begin{figure}[h] 
	\centering 
	\includegraphics[width=0.25\textwidth]{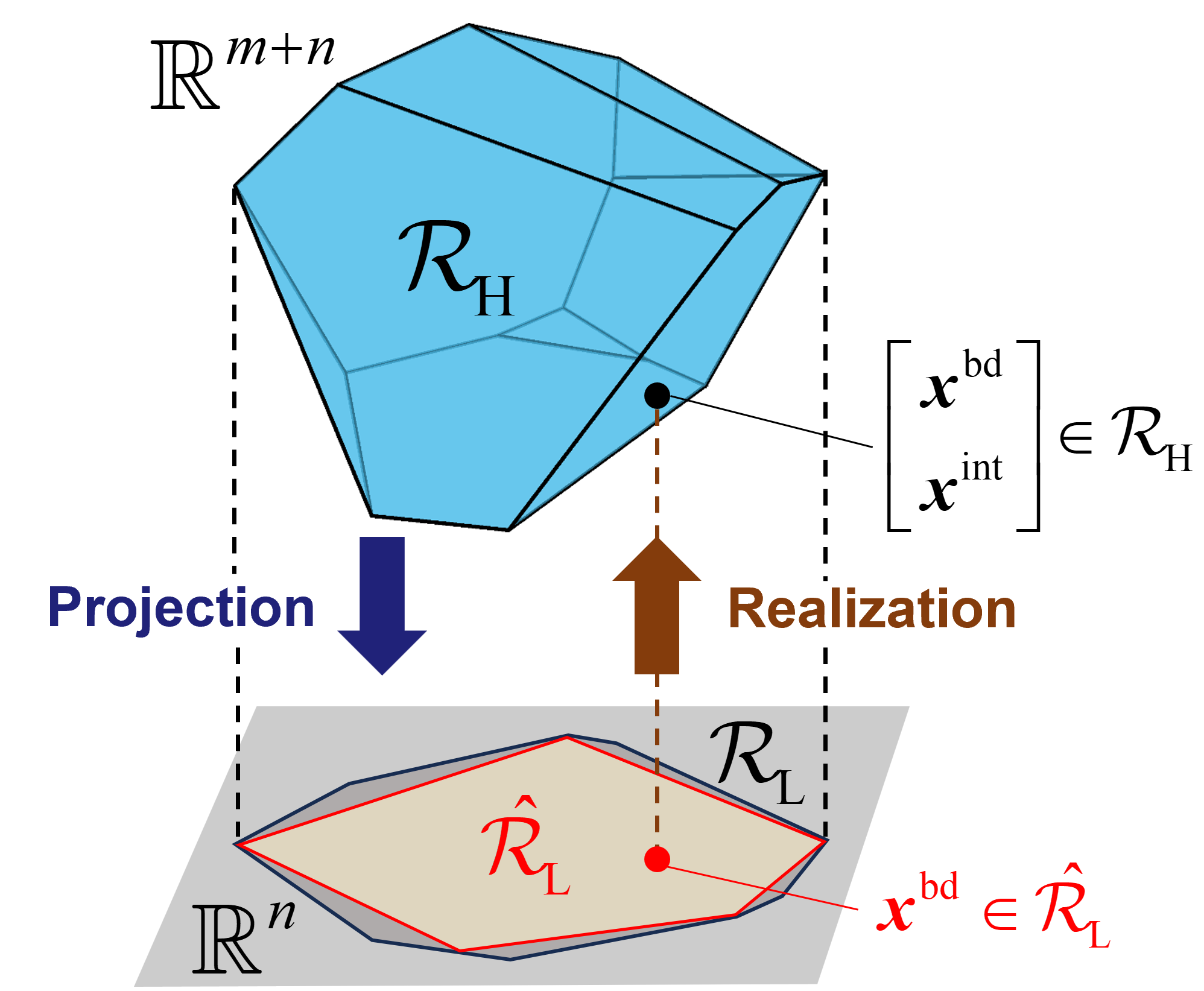}
	\caption{Schematic diagram of the inner-approximate the projected flexibility region.} 
	\label{fig-polyProjectionARO}
\end{figure}

The commonly used approaches are to calculate the inner-approximated the projected flexibility region $\mathcal{R}_\text{L}$ \cite{chenLeveragingTwoStageAdaptive2021,cuiNetworkCognizantTimeCoupledAggregate2021, wangAggregateFlexibilityVirtual2021,linHighdimensionTielineSecurity2023,wenAggregateFeasibleRegion2022,tanNonIterativeSolutionCoordinated2024,wenImprovedInnerApproximation2024}. 
Denote the inner-approximated aggregated flexibility region as $\hat{\mathcal{R}_\text{L}}$, such that $\hat{\mathcal{R}_\text{L}} \subseteq \mathcal{R}_\text{L}$. 
For every dispatch order of the boundary variables $\bm{x}^\text{bd}$ in $\hat{\mathcal{R}_\text{L}}$, there exists at least one realization of the internal variables $\bm{x}^\text{int}$ that satisfies the system's operation constraints, as shown in Fig.~\ref{fig-polyProjectionARO}

\subsection{Overview of Geometric-Based Methods}
\label{sec-shape_template}

Using the physical properties and geometric interpretations inherent in flexibility aggregation, shape template-based methods can effectively approximate the flexibility region $\mathcal{R}_\text{L}$~\cite{wangAggregateFlexibilityVirtual2021}. 
The key physical properties utilized in this context are the flexibility characteristics of the constituent internal devices. 
Recent geometry-based methods typically employ polytopal and ellipsoidal templates, derived from physical property analysis, to inner-approximate the aggregated flexibility region $\mathcal{R}_\text{L}$.

Since the system-level aggregated flexibility emerges from the collective behavior of all internal devices, if their individual flexibility regions share certain structural properties, the system's flexibility region can be approximated using a polytopal shape template defined by explicit linear constraints.
For instance, if all internal devices of an energy system exhibit energy storage property, such as energy storage devices, electric vehicles, and thermal controllable loads, the system's flexibility can be aggregated into an equivalent energy storage model. That is,

\begin{subequations} \label{eq-equ_storage}
	\begin{align} 
		& \underline{p}_{t} \leq p_{t} \leq \overline{p}_{t}, \forall 1 \leq t \leq T \\
		& \frac{\underline{E}_{t_1,t_2}}{\Delta t} \leq \sum_{\tau=t_1}^{t_2}p_{\tau} \leq \frac{\overline{E}_{t_1,t_2}}{\Delta t}, \forall 1 \leq t_1 < t_2 \leq T
	\end{align}
\end{subequations}

The shape template for the inner-approximated region can be expressed in a compact matrix form as follows:

\begin{equation} \label{eq-polytope_template_region}
	\hat{\mathcal{R}}_\text{L}(\bm{b})=
	\left\{
		\bm{x}^\text{bd} 
		\left\lvert\,
			\bm{A} \bm{x}^\text{bd} \leq \bm{b}
		\right.
	\right\}  
\end{equation} 

\noindent where the boundary variable vector $\bm{x}^\text{bd}$ is composed of $p_{t}$ across all time slots in this case.  
The matrix $\bm{A}$ is the parameters that correspond to the constraints \eqref{eq-equ_storage}, expressed in the storage form. 
The vector $\bm{b}$ represents the parameters of the inner-approximated region $\hat{\mathcal{R}}_\text{L}$, composed of all the paremeters $\underline{p}_{t}$, $\overline{p}_{t}$, $\nicefrac{\underline{E}_{t_1,t_2}}{\Delta t}$, and $\nicefrac{\overline{E}_{t_1,t_2}}{\Delta t}$ in this case.
The parameters in $\bm{b}$ need to be adjusted to ensure that condition $\hat{\mathcal{R}}_\text{L}(\bm{b}) \subseteq \mathcal{R}_\text{L}$ is satisfied. This adjustment can be achieved using our previously developed polytope-bound shrinking algorithm, as detailed in \cite{wangAggregateFlexibilityVirtual2021}.

\begin{figure}[h] 
	\centering 
	\includegraphics[width=0.4\textwidth]{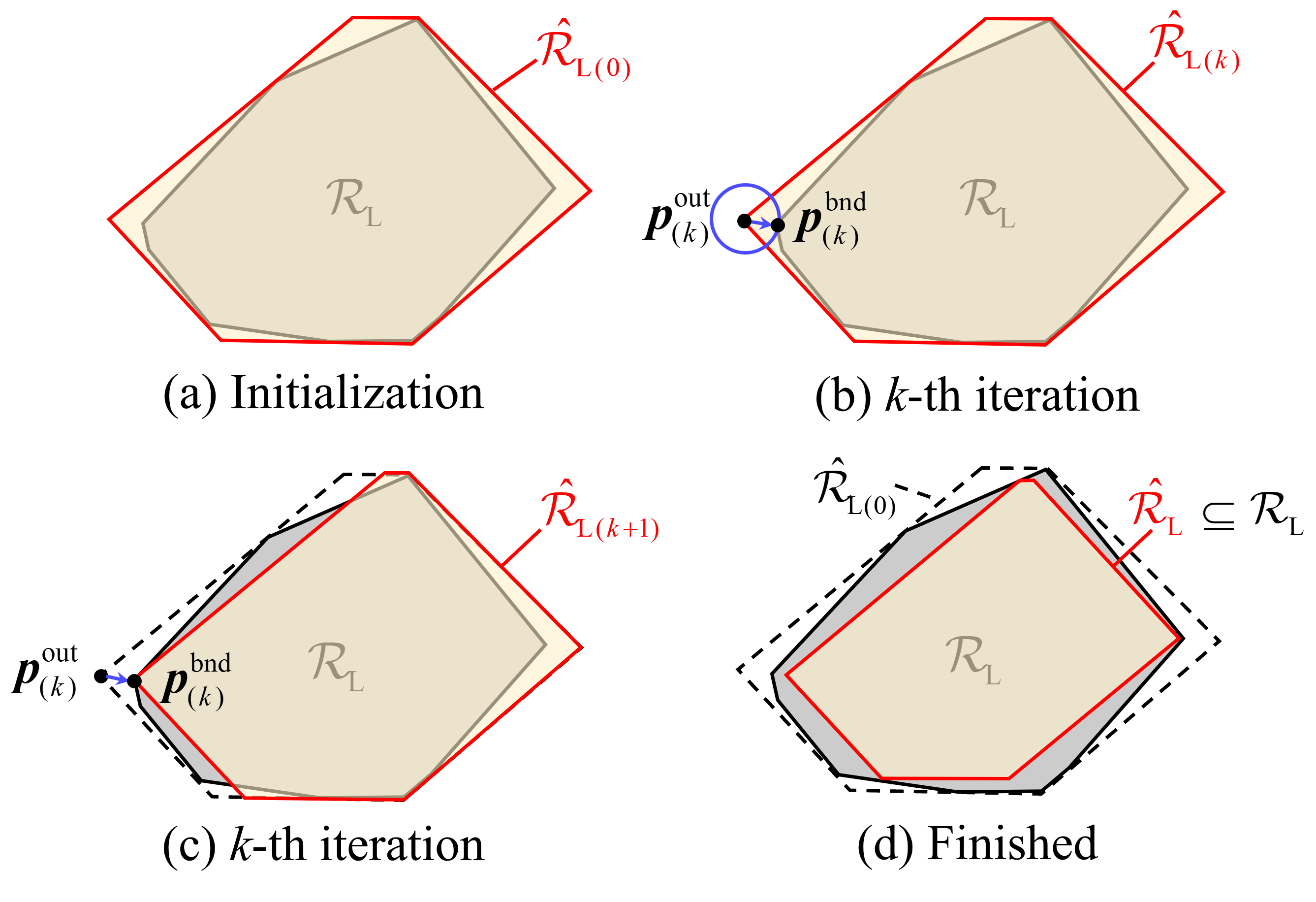}
	\caption{Schematic diagram of the polytope-bound shrinking algorithm: (a) Calculate the circumscribed polytope  $\hat{\mathcal{R}}_{\text{L}(0)}$ at the beginning of iteration; (b) Identify the outlier point $\bm{p}^\text{out}_{(k)}$ of the shrinking polytope $\hat{\mathcal{R}}_{\text{L}(k)}$ in the current $k$-th iteration. The nearest point on the  boundary of the projected polytope $\mathcal{R}_\text{L}$ is denoted as $\bm{p}^\text{bnd}_{(k)}$; (c) Shrink the boundaries to draw back the extreme point $\bm{p}^\text{out}_{(k)}$ to the boundary point $\bm{p}^\text{bnd}_{(k)}$ by adjusting parameters $\bm{b}_{(k)}$; (d) Repeat the bound shrinking process to finally obtain the inner-approximated polytope $\hat{\mathcal{R}}_\text{L} \subseteq \mathcal{R}_\text{L}$.}
	\label{fig-boundShrinkAlgorithm}
\end{figure}

Fig.~\ref{fig-boundShrinkAlgorithm} is an illustrative diagram that explains the procedure of the algorithm, with the detailed algorithms available in \cite{wangAggregateFlexibilityVirtual2021}. 
Inspired by geometric perspective, this algorithm starts from a circumscribed polytope of $\mathcal{R}_\text{L}$ with the given template shape and gradually shrinks the boundaries of the inner-approximated region $\hat{\mathcal{R}}_\text{L}(\bm{b}_{(k)})$ by gradually adjusting the values of parameters $\bm{b}_{(k)}$ in the $k$-th iteration.
Furthermore, if all nonzero elements in each row of the matrix $\bm{A}$ in \eqref{eq-polytope_template_region} have the same sign, the modified bound shrinking method \cite{wenImprovedInnerApproximation2024} can be applied to accelerate calculations.

When the flexibility region lacks an explicit constraint expression like \eqref{eq-polytope_template_region}, due to the presence of complex internal device variables or intricate coupling among boundary variables, it can be inner-approximated using a high-dimensional ellipsoid shape template \cite{zhenComputingMaximumVolume2018}, as illustrated in \eqref{eq-ellipsoid_template_region}.

\begin{equation} \label{eq-ellipsoid_template_region}
	\hat{\mathcal{R}}_\text{L}\left(\bm{E}, \bm{e}\right)= 
	\left\{
		\bm{x}^\text{bd} 
		\left\lvert\,
		\left\|
			\left(\bm{E}\right)^{-1}
			\left(\bm{x}^\text{bd}-\bm{e}\right)
		\right\| \leq 1
		\right.
	\right\}
\end{equation}

\noindent where $\bm{E} \in \mathbb{R}^{n \times n}$ and $\bm{e} \in \mathbb{R}^{n}$ represent the parameters of the inner-approximated ellipsoid feasible region. 
The matrix $\bm{E}$ governs the shape of the ellipsoid, and the vector $\bm{e}$ specifies its center.
This ellipsoidal shape template can self-adjust to better fit the shape of the projected region with fewer parameters.

To achieve the best possible approximation, the volume of the high-dimensional ellipsoid should be maximized.
This process is analogous to inflating an ellipsoid like a balloon within the projected region until it touches the boundaries and reaches its maximum volume. 
It can be expressed as the following robust optimization problem:

\begin{equation} \label{eq-robust_expression_for_ellipsoid}
	\max _{\bm{E} \succeq 0, \bm{e}, \bm{x}^\text{int}}
	\left\{
		\log \operatorname{det}\left(\bm{E}\right)
		\left\lvert
		\begin{array}{l}
				\forall \|\bm{\xi}\| \leq 1, \exists \bm{x}^\text{int}:\\
				\left[
					\begin{array}{c}
					\bm{E} \bm{\xi}+\bm{e} \\
					\bm{x}^\text{int}
					\end{array}
				\right] \in \mathcal{R}_\text{H}
		\end{array}
		\right.
	\right\}
\end{equation}

\noindent where $\bm{x}^\text{bd} := \bm{E} \bm{\xi} + \bm{e}$. Due to the difficulty in solving \eqref{eq-robust_expression_for_ellipsoid} in high-dimensional space, the problem can be simplified by introducing specific decision policies between variables $\bm{\xi}$ and $\bm{x}^\text{int}$. 
In engineering applications, adding linear or quadratic policies to the robust problem \eqref{eq-robust_expression_for_ellipsoid} is a common practice, which allows for an equivalent reformulation as a semi-definite programming (SDP) problem. 
Details on calculating the parameters $\bm{E}$ and $\bm{e}$ using an equivalent SDP can be found in \cite{zhenComputingMaximumVolume2018}.

In summary, geometric-based methods commonly employ polytopal and ellipsoidal shape templates to inner-approximate the aggregated flexibility region $\mathcal{R}_\text{L}$.
A template for polytopal shapes that match the flexibility properties can more accurately capture the boundaries of the actual flexibility region, thereby reducing conservativeness.
On the other hand, ellipsoidal templates are well-suited for representing irregular flexibility regions with few parameters.
However, their smooth and symmetric nature may result in more conservative approximations.
In practical applications, the choice between these templates can be guided by the availability of explicit constraints and the characteristics of the constraints within the system.

\section{Mathematical Model of a Regional Power Grid}
\label{sec-RPGMathModel}

This section introduces the mathematical model of an RPG and the operation constraints that make up the feasible region $\mathcal{R}_\text{H}$.
The network constraints with the DC power flow model, power balance constraints, $N$-1 security constraints, and generation unit constraints are considered.
For clarity in presentation, subscripts indicating the index of the RPG are omitted in this section.

\subsection{Network and Power Balance Constraints}

The operational constraints in a regional power grid are as follows.

\begin{subequations} \label{eq-RPGCons}
	\begin{align} 
		& \bm{p}^\text{inj}_t = \bm{A}^\text{tpg} \bm{p}_{t}^\text{tpg} + \bm{A}^{\text{rgu}} \bm{p}_{t}^{\text{rgu}} - \bm{A}^\text{tie} \bm{p}_{t}^\text{tie} - \bm{p}_{t}^{\text{ld}} \label{eq-RPGBusInj} \\
		& -\overline{\bm{p}}^{\text{line}} \leq \bm{S}^\text{PTDF}\bm{p}^\text{inj}_t \leq \overline{\bm{p}}^{\text{line}} \label{eq-RPGLineLimit} \\
		& \bm{\theta}_{t} = \theta_{t}^{\text{ref}} + \bm{B}^{-1}\bm{p}^\text{inj}_t, \bm{\theta}_{t}^\text{bd}=
		\bm{M}^\text{bd} \bm{\theta}_{t} \label{eq-RPGAngle} 
	\end{align}
\end{subequations}

\noindent where \eqref{eq-RPGBusInj} calculates the net injection power $\bm{p}^\text{inj}_t$ of all buses at time $t$. 
$\bm{p}^\text{tpg}_t$, $\bm{p}^\text{rgu}_t$ are vectors composed of thermal power generators and renewable generators' output power at time $t$.
$\bm{p}^\text{tie}_t$ denotes the vector composed of tie-lines' output power at time $t$.
$\bm{p}^\text{ld}_t$ denotes the vector composed of buses' load power at time $t$.
The matrices $\bm{A}^\text{tpg}$, $\bm{A}^\text{rgu}$, and $\bm{A}^\text{tie}$ represent device-bus associations, mapping the indices of each respective device to the corresponding bus indices.
\eqref{eq-RPGLineLimit} is the transmission line power capability constraints. 
$\bm{S}^\text{PTDF}$ denotes the power transfer distribution factor (PTDF) matrix.
The voltage angles of the boundary buses that are directly connected to AC tie-lines, denoted as $\bm{\theta}_{t}^\text{bd}$, are calculated in \eqref{eq-RPGAngle}. 
$\theta_{t}^{\text{ref}}$ denotes the voltage phase angle of the reference node in this RPG relative to the global reference node.
$\bm{B}$ denotes the susceptance matrix.
$\bm{M}^\text{bd}$ is used to select all the boundary buses indices that are directly connected to AC tie-lines.

\textit{Remark:} The regional power grid is modeled using the DC power flow formulation in \eqref{eq-RPGCons}, with a focus solely on active power variables. 
To further incorporate reactive power, the model can be generalized to a high-precision linearized AC power flow framework using the methods proposed in \cite{yangStateIndependentLinearPower2017, yangNovelNetworkModel2017}. 
The flexibility aggregation method proposed herein remains valid as long as linear relationships between variables are maintained.

The power balance equation within the RPG is as follows:

\begin{equation} \label{eq-RPGPowerBalance}
	\bm{1}^{\top} \bm{p}_{t}^\text{tpg} + \bm{1}^{\top} \bm{p}_{t}^{\text{rgu}} = \bm{1}^{\top} \bm{p}_{t}^\text{tie} + \bm{1}^{\top} \bm{p}_{t}^{\text{ld}}
\end{equation}

\subsection{Generation Units Constraints}

The operational constraints for all renewable generation station $i \in \mathcal{I}^\text{rgu}$ are

\begin{equation} \label{eq-RPGRgCons}
	\underline{p}^{\text{rgu}}_{i} \leq p_{i,t}^{\text{rgu}} \leq \overline{p}^{\text{fore}}_{i,t}
\end{equation}

\noindent where $\overline{p}^{\text{fore}}_{i,t}$ denotes the maximum predicted output power of the $i$-th renewable generation station at time $t$.

Furthermore, the operational constraints for all thermal power generators $i \in \mathcal{I}^\text{tpg}$ are as follows:

\begin{subequations} \label{eq-RPGGenCons}
	\begin{align} 
		& \underline{p}^\text{tpg}_{i} + Rd^\text{tpg}_{i,t} \leq p_{i,t}^\text{tpg} \leq \overline{p}^\text{tpg}_{i} - Ru^\text{tpg}_{i,t} \label{eq-RPGGenPower} \\ 
		& \underline{r}^\text{tpg}_{i,t} \leq \left(p_{i,t+1}^\text{tpg}-p_{i,t}^\text{tpg}\right) /{\Delta t} \leq \overline{r}^\text{tpg}_{i,t} \label{eq-RPGGenRampRate} \\
		& 0 \leq Rd^\text{tpg}_{i,t} \leq \overline{Rd}^\text{tpg}_{i,t}, 0 \leq Ru^\text{tpg}_{i,t} \leq \overline{Ru}^\text{tpg}_{i,t} \label{eq-RPGGenReserve} \\
		& \sum\limits_{i \in \mathcal{I}^\text{tpg}} Rd^\text{tpg}_{i,t} \geq Rd^\text{req}_{t},  \sum\limits_{i \in \mathcal{I}^\text{tpg}} Ru^\text{tpg}_{i,t} \geq Ru^\text{req}_{t} \label{eq-RPGGenReserveTotal}
	\end{align} 
\end{subequations}

\noindent where \eqref{eq-RPGGenPower} and \eqref{eq-RPGGenRampRate} are the power and ramp rate constraints, respectively. 
\eqref{eq-RPGGenReserve} and \eqref{eq-RPGGenReserveTotal} are the reserve constraints.
$Rd^\text{tpg}_{i,t}, Ru^\text{tpg}_{i,t}$ denote down and up reserve of $i$-th thermal power generator at time $t$.
$Rd^\text{req}_{t}$ and $Ru^\text{req}_{t}$ denote the total down and up reserve requirements of the entire RPG at time $t$, determined by the dispatch center based on the fluctuation in the generation of renewable energy and load power \cite{yangAnalyticalReformulationStochastic2020}.

\subsection{N-1 Security Constraints}

To safeguard the power grid when a fault occurs in critical equipment, such as transmission lines and generators, the $N$-1 security constraints are considered in the security constants of RPG. For $\forall c \in \mathcal{C}$:

\begin{subequations} \label{eq-RPGN1Cons}
	\begin{align} 
		& \bm{p}^\text{inj}_{t,c} = \bm{A}^\text{tpg} \bm{p}_{t,c}^\text{tpg} + \bm{A}^{\text{rgu}} \bm{p}_{t,c}^{\text{rgu}} - \bm{A}^\text{tie} \bm{p}_{t}^\text{tie} - \bm{p}_{t}^{\text{ld}} \\
		& -\overline{\bm{p}}^{\text{line}} \leq \bm{S}^\text{PTDF}\bm{p}^\text{inj}_{t,c} \leq \overline{\bm{p}}^{\text{line}} \\
		& \bm{1}^{\top} \bm{p}_{t,c}^\text{tpg} + \bm{1}^{\top} \bm{p}_{t,c}^{\text{rgu}} = \bm{1}^{\top} \bm{p}_{t}^\text{tie} + \bm{1}^{\top} \bm{p}_{t}^{\text{ld}} \\
		& \underline{\bm{p}}^\text{tpg}  \leq \bm{p}_{t,c}^\text{tpg} \leq \overline{\bm{p}}^\text{tpg}, \underline{\bm{p}}^{\text{rgu}} \leq \bm{p}_{t,c}^{\text{rgu}} \leq \overline{\bm{p}}^{\text{fore}}_{t} \\ 
		& \left| \bm{p}_{t,c}^\text{tpg} - \bm{p}_{t}^\text{tpg} \right| \leq \Delta \bm{p}^\text{tpg}, \left| \bm{p}_{t,c}^\text{rgu} - \bm{p}_{t}^\text{rgu} \right| \leq \Delta \bm{p}^\text{rgu} \\
		& \bm{\theta}_{t,c} = \theta_{t}^{\text{ref}} + \bm{B}^{-1}_c \bm{p}^\text{inj}_{t,c}, \left| \bm{\theta}_{t,c} - \bm{\theta}_{t} \right| \leq \Delta \bm{\theta}
	\end{align}
\end{subequations}

\noindent where the subscript $c$ represents the index of $N$-1 contingency scenarios, and $\mathcal{C}$ is the index set of key contingency. $\Delta \bm{p}^\text{tpg}$, $\Delta \bm{p}^\text{rgu}$ and $\Delta \bm{\theta}$ denote the maximum allowable variations of the power output of thermal power generators and renewable generation stations, as well as the voltage angles of the buses in the contingency scenarios, respectively.

\subsection{Stochastic Factors Consideration}

The forecasts for the load and renewable energy generation contain inherent random errors, which introduce stochastic factors into the RPGs model. 
Using historical data, existing methods can convert the constraints with stochastic variables into deterministic constraints.
For example, the Gaussian Mixture Model (GMM) approximates the probability distribution of stochastic variables, enabling the conversion of constraints into deterministic equivalents through the chance constraint method at a specified confidence level \cite{yangAnalyticalReformulationStochastic2020,wangStochasticFlexibilityEvaluation2024}.
Moreover, distributionally robust optimization converts joint chance constraints into deterministic constraints by considering the distance between the potential variable distributions and the empirical distribution \cite{ordoudisEnergyReserveDispatch2021}.
Using either approach, the stochastic RPG model can be effectively transformed into a deterministic model, which serves as the model basis for computing the aggregated RPG model proposed in this work.

\subsection{Model Integration}

By integrating all the operational constraints of RPG \eqref{eq-RPGCons}-\eqref{eq-RPGN1Cons}, the feasible operation region of the RPG can be defined in a high-dimensional polytopal form as follows:

\begin{equation} 
	\mathcal{R}_{\text{H}}=
	\left\{
		\left[
			\begin{array}{l}
				\bm{x}^\text{bd} \\
				\bm{x}^\text{int}
			\end{array} 
		\right]
		\left\lvert\,
			\begin{array}{l}
				\bm{x}^\text{bd} = \bm{C} \bm{x}^\text{int} + \bm{d}  \\
				\bm{F} \bm{x}^\text{int} \leq \bm{f}
			\end{array} 
		\right.
	\right\}
\end{equation}

\noindent where the matrices $\bm{C}$, $\bm{D}$, $\bm{F}$, vectors $\bm{d}$ and $\bm{f}$ are constant parameters that can be obtained by integrating constraints \eqref{eq-RPGCons}-\eqref{eq-RPGN1Cons}. 
$\bm{x}^\text{int}$ and $\bm{x}^\text{bd}$ are internal and boundary variables, defined as follows.

The internal variables $\bm{x}^\text{int}$ include all the variables of power generation units and network states, under normal and $N$-1 contingency operation conditions.

The boundary variables $\bm{x}^\text{bd}$ include the output power of the tie-lines $\bm{p}^\text{tie}_t$, and the voltage angles of the boundary buses connected directly to the AC tie-lines $\bm{\theta}^\text{bd}_t$, as follows:

\begin{equation} \label{eq-DefBdVar}
	\bm{x}^\text{bd} := \underset{t \in \mathcal{T}}{\operatorname{col}}(\bm{x}^\text{bd}_t),
	\bm{x}^\text{bd}_t :=\left[
		\left(\bm{p}^\text{tie}_t\right)^{\top}
		\left(\bm{\theta}^\text{bd}_t\right)^{\top}
		\right]^{\top}
\end{equation}

This choice of boundary variables is based on the fact that the transmission power of the DC tie-lines is directly controlled by converters, while the AC tie-line power transmission depends on the voltage angles of the boundary buses \cite{linHighdimensionTielineSecurity2023}, governed by power balance and network constraints.

\begin{equation} \label{eq-RPGTieLine}
	p_{r, ij, t}^\text{tie}=
	-p_{s, ji, t}^\text{tie}=
	\left(\theta_{r, i, t}^\text{bd}-\theta_{s, j, t}^\text{bd}\right) / {x_{ij}}
\end{equation}

\noindent where $\theta_{r, i, t}^\text{bd}$ and $\theta_{s, j, t}^\text{bd}$ denote the voltage angles of the $i$-th bus in region $r$ and the $j$-th bus in region $s$ at time $t$, respectively. 
$p_{r, ij, t}^\text{tie}$ and $p_{s, ji, t}^\text{tie}$ represent power flows from region $r$ to region $s$ through the tie-line $ij$ and vice versa, respectively, and $x_{ij}$ is the reactance of this AC tie-line  $ij$.

Subsequently, the flexibility aggregation problem of an RPG can be converted to the projection problem of the high-dimensional feasibility region $\mathcal{R}_{\text{H}}$ onto the lower-dimensional region $\mathcal{R}_\text{L} \subset \operatorname{space}(\bm{x}^\text{bd})$.


\section{Aggregated Model Calculation for A Regional Power Grid}
\label{sec-RPGAgg}

\subsection{Dimension-Decomposition-Based Calculation of Aggregated Flexibility Region}

\begin{figure}[h] 
	\centering 
	\includegraphics[width=0.48\textwidth]{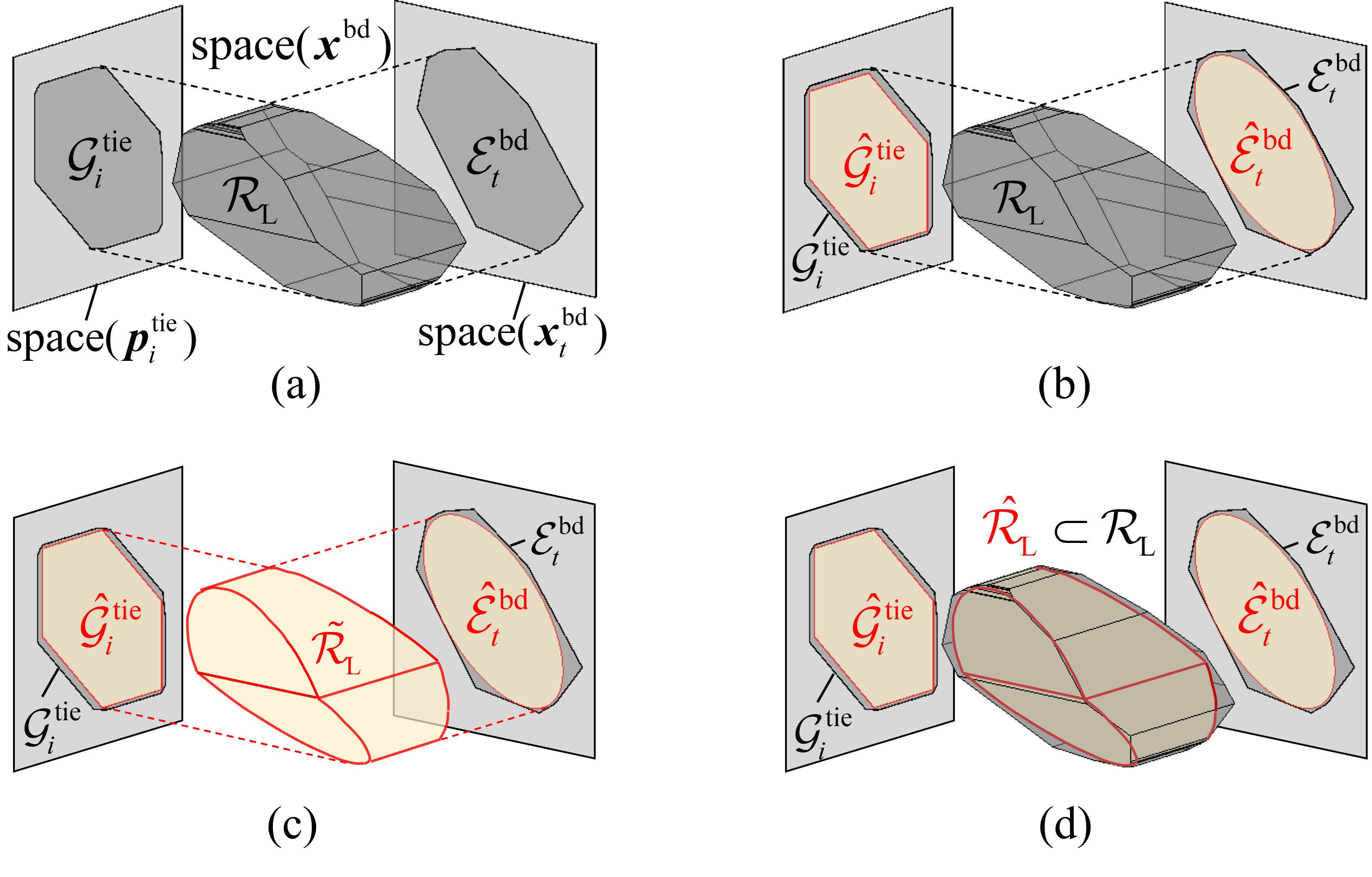}
	\caption{Dimension decomposition and recombination process of the aggregated flexibility region: (a) Project the aggregated flexibility region $\mathcal{R}_\text{L}$ onto two groups of lower-dimensional subspaces; (b) Inner-approximate the polytopes in the subspaces with appropriate shape templates; (c) Recombine the inner-approximated polytopes from subspaces to obtain the aggregated flexibility region $\tilde{\mathcal{R}}_\text{L}$ in $\operatorname{space}(\bm{x}^\text{bd})$; (d) Adjust the parameters to obtain the inner-approximated aggregated flexibility region $\hat{\mathcal{R}}_\text{L} \subseteq \mathcal{R}_\text{L}$.} 
	\label{fig-dimDecompose}
\end{figure}

This section presents the proposed dimension-decomposition-based method for calculating the aggregated flexibility region of an RPG. 
There are four main steps in the process.
An illustrative diagram of each step is provided in Fig.~\ref{fig-dimDecompose}. 

\begin{figure}[h] 
	\centering 
	\includegraphics[width=0.18\textwidth]{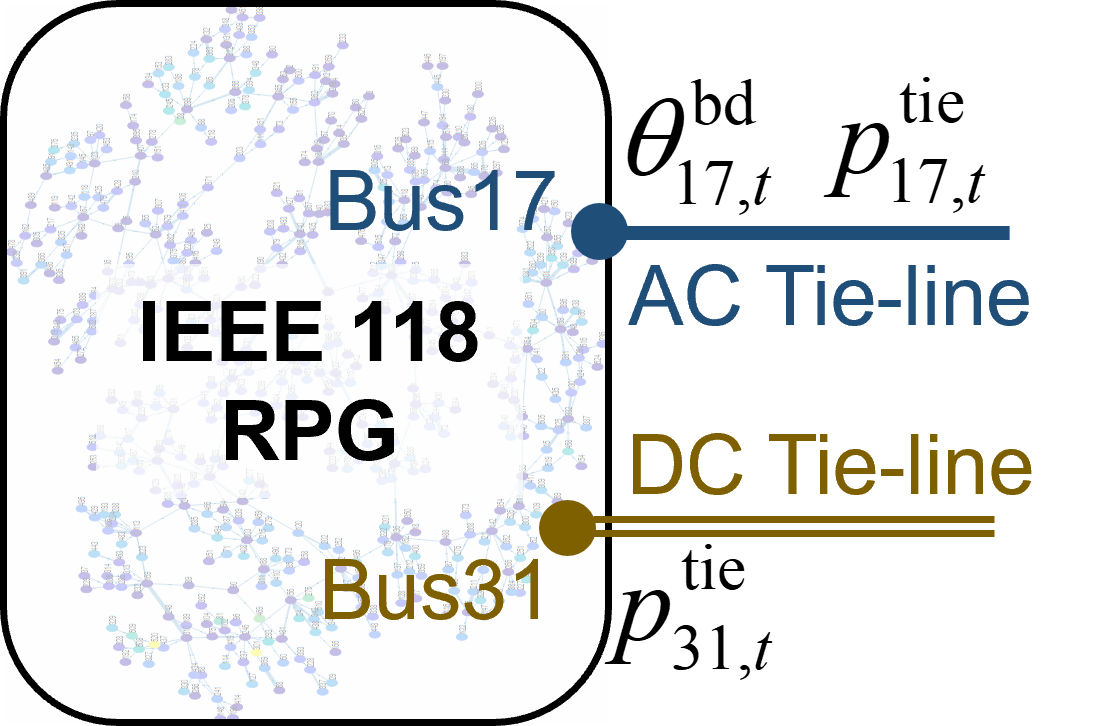}
	\caption{Schematic diagram of the IEEE 118 regional power grid.} 
	\label{fig-case118schematic}
\end{figure}

To illustrate the method, we use an IEEE 118-bus RPG example, introduced step by step alongside the methodology.
This system includes two tie-lines: an AC tie-line connected to Bus 17 and a DC tie-line connected to Bus 31.
The schematic diagram of this system is depicted in Fig.~\ref{fig-case118schematic}.

\subsubsection{Analysis of Boundary Variables}
\label{sec-RPGCoupling}

\begin{figure}[h] 
	\centering 
	\includegraphics[width=0.4\textwidth]{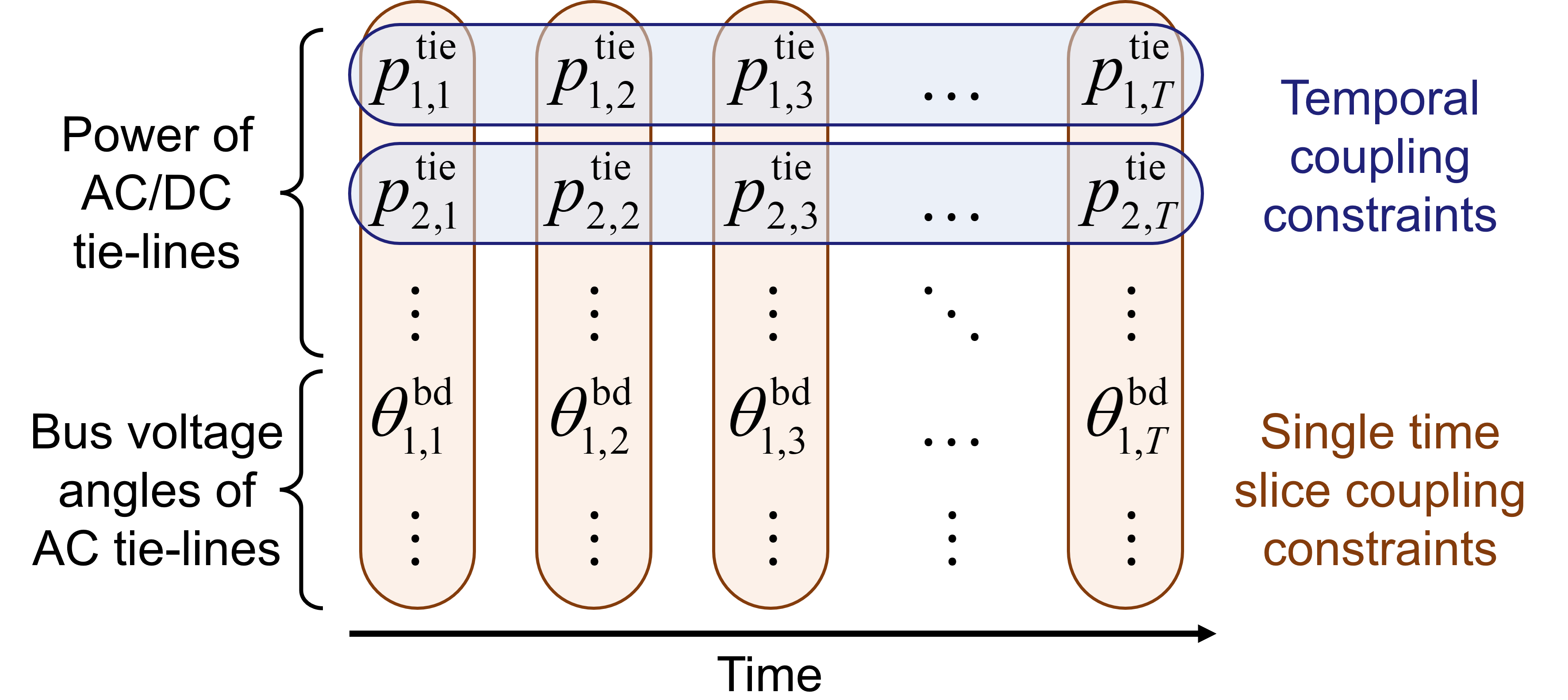}
	\caption{The coupling constraints of boundary variables for a regional power grid and two categories of constraints.} 
	\label{fig-boundaryVariablesCouplingRelationship}
\end{figure}

To calculate the aggregated flexibility region $\mathcal{R}_\text{L}$ of an RPG in $\operatorname{space}(\bm{x}^\text{bd})$, it is crucial to understand the coupling relationships between the boundary variables $\bm{x}^\text{bd}$.
By analyzing the coupling constraints of the boundary variables in \eqref{eq-RPGCons}-\eqref{eq-RPGN1Cons}, all the constraints can be categorized into two types. Consequently, all the boundary variables can be categorized into two categories of variable sets, as illustrated in Fig.~\ref{fig-boundaryVariablesCouplingRelationship}.

The first category, temporal coupling constraints, addresses the active output power constraints across all time slots for a single tie-line. 
The corresponding variables in each group are denoted as 

\begin{equation} \label{eq-DefPtie}
	\bm{p}^\text{tie}_i := \underset{t \in \mathcal{T}}{\operatorname{col}} (\bm{p}^\text{tie}_{i,t}) := \bm{S}^\text{bd}_i \bm{x}^\text{bd}, \forall i \in \mathcal{I}^\text{tie}
\end{equation}

\noindent This form of coupling is mainly driven by generator operational constraints \eqref{eq-RPGGenCons} and power balance constraints \eqref{eq-RPGPowerBalance}. 
The active output power across time slots for a single tie-line exhibits temporal coupling characteristics, similar to those observed in thermal power generators, which are regulated by both power and ramp rate constraints.

The second category, single time slice coupling constraints, encompasses coupling constraints for boundary variables within each individual time slice $t$. 
The corresponding variables in each group are denoted as

\begin{equation} \label{eq-DefXbdt}
	\bm{x}^\text{bd}_t := \bm{H}^\text{bd}_t \bm{x}^\text{bd}, \forall t \in \mathcal{T}
\end{equation}

\noindent They are governed by constraints that apply independently within each time slice, such as network constants \eqref{eq-RPGLineLimit}, power balance \eqref{eq-RPGPowerBalance}, and $N$-1 security constraints \eqref{eq-RPGN1Cons}.

In terms of the IEEE 118 RPG, the boundary variables include the output power of two tie-lines ($p^\text{tie}_{17,t}$, $p^\text{tie}_{31,t}$) and the voltage angle of Bus 17 ($\theta^\text{bd}_{17,t}$) across all time slots $t \in \mathcal{T}$.
Accordingly, the first category includes the output power constraints of each tie-line across all time slots. 
The corresponding two groups of variables are $\bm{p}^\text{tie}_{17} := \operatorname{col}_{t \in \mathcal{T}}(p^\text{tie}_{17,t})$ and $\bm{p}^\text{tie}_{31} := \operatorname{col}_{t \in \mathcal{T}}(p^\text{tie}_{31,t})$, respectively.

\noindent The second category includes the constraints of the output power of two tie-lines and the voltage angle of Bus 17 within each time slot $t \in \mathcal{T}$. 
The corresponding $T$ groups of variables are $\bm{x}^\text{bd}_t := \left[p^\text{tie}_{17,t}, p^\text{tie}_{31,t}, \theta^\text{bd}_{17,t}\right]^\top$.

\subsubsection{Dimension Decomposition Projection and Inner-Approximation}

Based on the categorization results of all the boundary variables in the previous step, all the coupling relationships of the boundary variables in an RPG can be divided into two categories accordingly.
From a geometric perspective, the aggregated flexibility region $\mathcal{R}_\text{L} \subset \operatorname{space}(\bm{x}^\text{bd})$ can be decomposed into two corresponding groups of subspaces: $\operatorname{space}(\bm{p}^\text{tie}_i)$ for $i \in \mathcal{I}^\text{tie}$ and $\operatorname{space}(\bm{x}^\text{bd}_t)$ for $t \in \mathcal{T}$.

To express the coupling relationships within these subspaces, the aggregated flexibility region $\mathcal{R}_\text{L}$ can be projected onto lower-dimensional subspaces. 
Denote the projected polytopes by $\mathcal{G}^\text{tie}_i \subset \operatorname{space}(\bm{p}^\text{tie}_i)$ for $i \in \mathcal{I}^\text{tie}$, and $\mathcal{E}^\text{bd}_t \subset \operatorname{space}(\bm{x}^\text{bd}_t)$ for $t \in \mathcal{T}$, respectively. An illustrated diagram is shown in Fig.~\ref{fig-dimDecompose}(a).
Subsequently, to obtain the explicit expression of the projected polytopes $\mathcal{E}^\text{bd}_t$ and $\mathcal{G}^\text{tie}_i$, we use the methods provided in Section \ref{sec-shape_template} to inner-approximate them, as demonstrated in Fig.~\ref{fig-dimDecompose}(b).

Since the temporal coupling constraints for a single tie-lie exhibit generator-like temporal coupling characteristics, the polytopes $\mathcal{G}^\text{tie}_i$ can be inner-approximated by the generator-like polytopal shape templates, denoted as $\hat{\mathcal{G}}^\text{tie}_i$.

\begin{equation}   \label{eq-RPGPolyInnerApprox}
	\hat{\mathcal{G}}^\text{tie}_i\left(\bm{b}^\text{tie}_i\right) := \left\{
		\bm{p}^\text{tie}_i 
		\left\lvert\,
		\begin{array}{c}
			\underline{r}^\text{tie}_{i,t} \leq (p^\text{tie}_{i,t+1} - p^\text{tie}_{i,t})/{\Delta t} \leq \overline{r}^\text{tie}_{i,t}, \\
			\underline{p}^\text{tie}_{i,t} \leq p^\text{tie}_{i,t} \leq \overline{p}^\text{tie}_{i,t},\forall t \in \mathcal{T}
		\end{array}
		\right.
	\right\}
\end{equation}

\noindent where the vector $\bm{b}^\text{tie}_i$ collects all the parameters of the generator-like polytope shape template, defined as follows:

\begin{equation} 
	\bm{b}^\text{tie}_i:= \\
	\left[
		(\overline{\bm{p}}^\text{tie}_{i})^{\top}
		-(\underline{\bm{p}}^\text{tie}_{i})^{\top}
		(\overline{\bm{r}}^\text{tie}_{i})^{\top}
		-(\underline{\bm{r}}^\text{tie}_{i})^{\top}
	\right]^{\top}
\end{equation}

\noindent where $\overline{\bm{p}}^\text{tie}_{i}:=\operatorname{col}_{t}(\overline{p}^\text{tie}_{i,t})$, $\underline{\bm{p}}^\text{tie}_{i}:=\operatorname{col}_{t}(\underline{p}^\text{tie}_{i,t})$, $\overline{\bm{r}}^\text{tie}_{i}:=\operatorname{col}_{t}(\overline{r}^\text{tie}_{i,t})$, $\underline{\bm{r}}^\text{tie}_{i}:=\operatorname{col}_{t}(\underline{r}^\text{tie}_{i,t})$. 
$\bm{b}^\text{tie}_i$ can be calculated based on Section \ref{sec-shape_template}. 
Then, \eqref{eq-RPGPolyInnerApprox} can be rewritten as the following compact matrix form:

\begin{equation}   \label{eq-RPGPolyMatrix}
	\hat{\mathcal{G}}^\text{tie}_i\left(\bm{b}^\text{tie}_i\right) = \left\{
		\bm{p}^\text{tie}_i 
		\left\lvert\,
			\bm{A}^\text{tie}_i \bm{p}^\text{tie}_i \leq \bm{b}^\text{tie}_i
		\right.
	\right\}
\end{equation}

The single time slice coupling constraints are governed by the complex coupling relationships among the coupling variables that lack an explicit expression; the polytopes $\mathcal{E}^\text{bd}_t$ can be inner-approximated by the ellipsoidal shape templates, denoted as $\hat{\mathcal{E}}^\text{bd}_t$. 

\begin{equation}  \label{eq-RPGEllipseInnerApprox}
	\hat{\mathcal{E}}_{t}^\text{bd}\left(\bm{E}_{t}^\text{bd}, \bm{e}_{t}^\text{bd}\right):=
	\left\{
		\bm{x}_{t}^\text{bd} 
		\left\lvert\,
			\left\|\left(\bm{E}_{t}^\text{bd}\right)^{-1}\left(\bm{x}_{t}^\text{bd}-\bm{e}_{t}^\text{bd}\right)\right\| \leq 1
		\right.
	\right\}
\end{equation}

\noindent The parameters $\bm{E}_{t}^\text{bd}$ and $\bm{e}_{t}^\text{bd}$ can be calculated based on Section \ref{sec-shape_template}.

In particular, the independence of projection and inner approximation calculations in each subspace allows the computation of parameters, $\bm{b}^\text{tie}_i$ for $i \in \mathcal{I}^\text{tie}$, as well as $\bm{E}_{t}^\text{bd}$ and $\bm{e}_{t}^\text{bd}$ for $t \in \mathcal{T}$, to be efficiently accelerated by parallel computing.

\begin{figure}[h] 
	\centering 
	\includegraphics[width=0.42\textwidth]{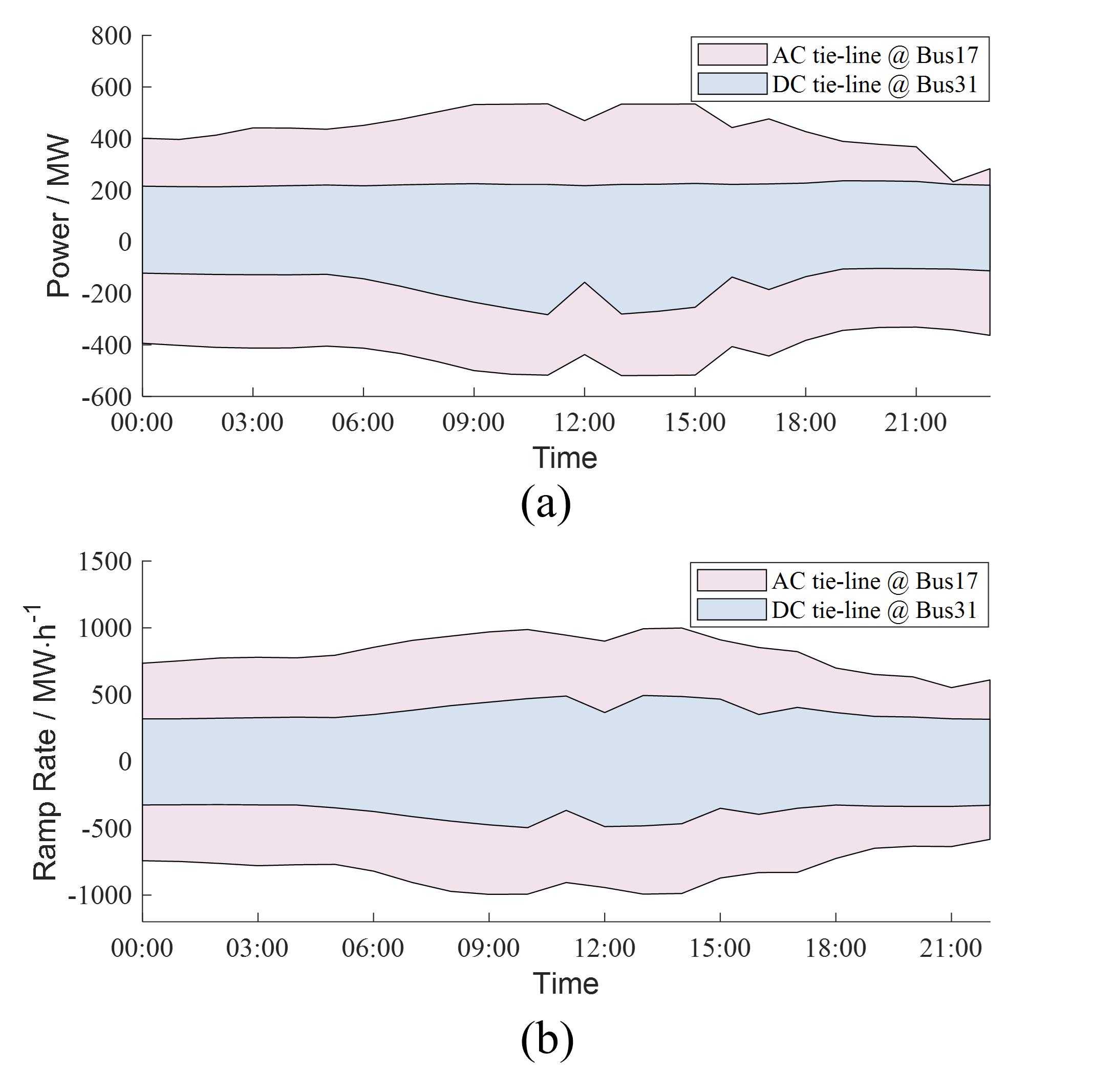}
	\caption{The aggregated parameters of two tie-lines in the IEEE 118 RPG: (a) Power bounds and (b) Ramp rate bounds.} 
	\label{fig-case118powerRamp}
\end{figure}

As for the IEEE 118 RPG case, the temporal coupling flexibility region across all time slots including the power bounds ($\overline{\bm{p}}^\text{tie}_{17}$, $\underline{\bm{p}}^\text{tie}_{17}$, $\overline{\bm{p}}^\text{tie}_{31}$, $\underline{\bm{p}}^\text{tie}_{31}$) and ramp rate bounds ($\overline{\bm{r}}^\text{tie}_{17}$, $\underline{\bm{r}}^\text{tie}_{17}$, $\overline{\bm{r}}^\text{tie}_{31}$, $\underline{\bm{r}}^\text{tie}_{31}$) of these two tie-lines are shown in Fig.~\ref{fig-case118powerRamp}.

\begin{figure}[h] 
	\centering 
	\includegraphics[width=0.4\textwidth]{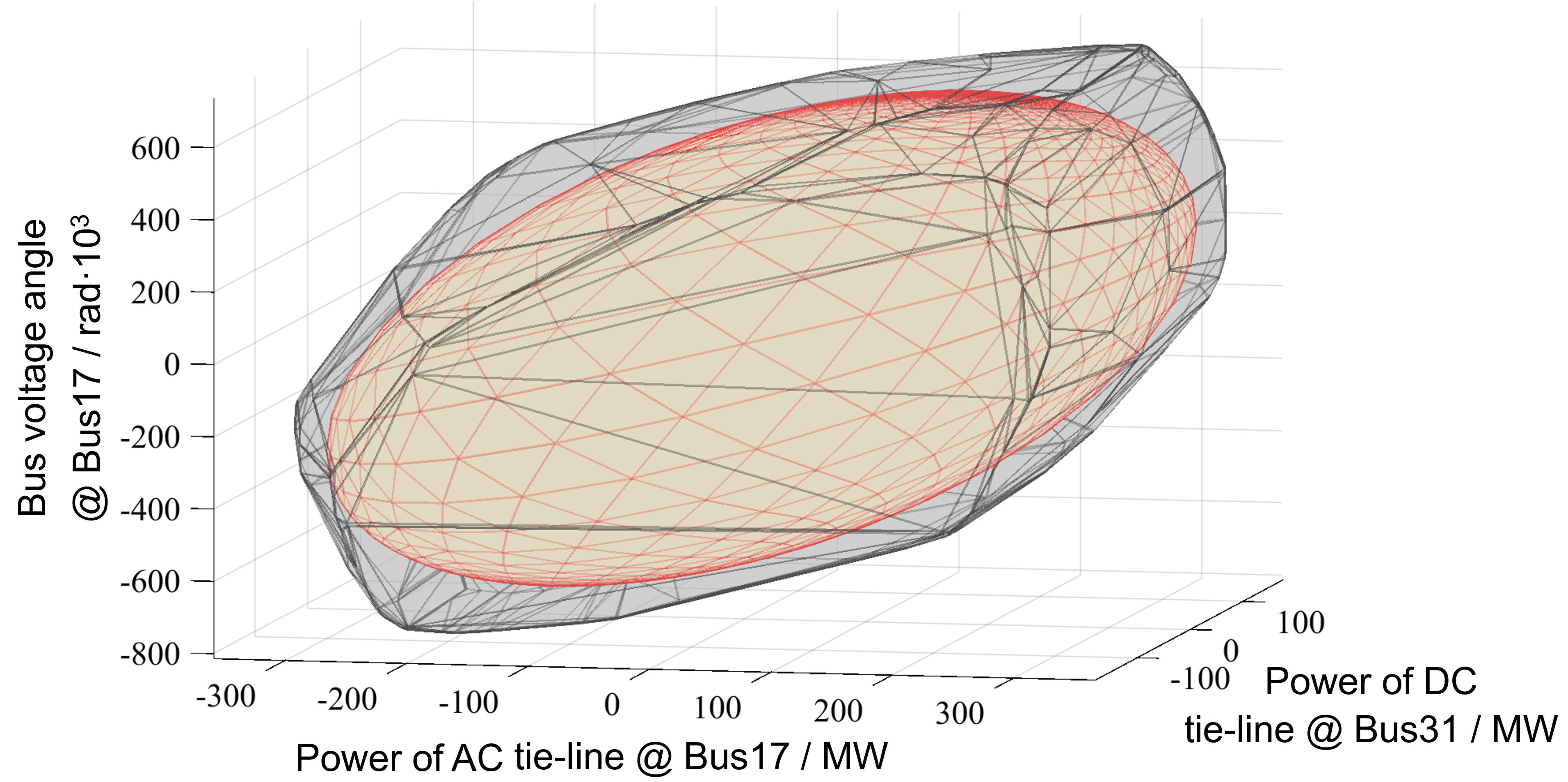}
	\caption{Coupling relationship among three boundary variables and inner-approximated flexibility region at time slice 12:00.} 
	\label{fig-timeSliceCoupling}
\end{figure}

In addition, taking the time slot $t_0=12:00$  as an example, the coupling relationship between the boundary variables $\bm{x}^\text{bd}_{t_0} := [p^\text{tie}_{17,t_0}, p^\text{tie}_{31,t_0}, \theta^\text{bd}_{17,t_0}]^\top$ is shown in Fig.~\ref{fig-timeSliceCoupling}.
The ellipsoid represents the inner-approximated flexibility region $\hat{\mathcal{E}}_{t_0}^\text{bd}$ with an ellipsoidal shape template.
The gray polytope represents the actual flexibility region $\mathcal{E}^\text{bd}_{t_0}$ obtained by searching for all extreme points.
In practice, the RPG may have far more tie-lines, making it intractable to find all the extreme points of the actual flexibility region like this case.
Therefore, the ellipsoidal shape template is a convenient tool to self-adaptively inner-approximate the flexibility region with fewer parameters.

\subsubsection{Dimension Recombination}

As demonstrated in Fig.~\ref{fig-dimDecompose}(c), to form the overall flexibility region, the inner-approximated regions $\mathcal{E}^\text{bd}_t$ for $t \in \mathcal{T}$, and $\mathcal{G}^\text{tie}_i$ for $i \in \mathcal{I}^\text{tie}$ are recombined in $\operatorname{space}(\bm{x}^\text{bd})$ as follows:

\begin{equation} \label{eq-RPGFlexRecombined}
	\tilde{\mathcal{R}}_\text{L}:=\left\{
		\bm{x}^\text{bd} 
		\left\lvert\,
		\begin{array}{c}
			\bm{S}^\text{tie}_i \bm{x}^\text{bd} \in \hat{\mathcal{G}}^\text{tie}_i\left(\tilde{\bm{b}}^\text{tie}_i\right),
			\forall i \in \mathcal{I}^\text{tie} \\
			\bm{H}^\text{bd}_t \bm{x}^\text{bd} \in \hat{\mathcal{E}}_{t}^\text{bd}\left(\tilde{\bm{E}}_{t}^\text{bd}, \bm{e}_{t}^\text{bd}\right),
			\forall t \in \mathcal{T}
		\end{array}
		\right.
	\right\}
\end{equation}

\noindent where the constant matrices $\bm{S}^\text{tie}_i$ and $\bm{H}^\text{bd}_t$ are used to select the elements $\bm{p}^\text{tie}_i$ and $\bm{x}^\text{bd}_t$ from all the boundary variables $\bm{x}^\text{bd}$ defined in \eqref{eq-DefPtie} and \eqref{eq-DefXbdt}, respectively.
The recombination region is an approximation of $\mathcal{R}_\text{L}$, denoted as $\tilde{\mathcal{R}}_\text{L}$, where the parameters $\tilde{\bm{b}}^\text{tie}_i$ and $\tilde{\bm{E}}_{t}^\text{bd}$ still need further adjustments in the following step.

\subsubsection{Parameter Adjustment}

To finally obtain the inner-approximated $\hat{\mathcal{R}}_\text{L} \subseteq \mathcal{R}_\text{L}$, further adjustments are required to the parameters of $\tilde{\mathcal{R}}_\text{L}$. 
Parameter adjustment can be achieved iteratively by shrinking the bounds of $\hat{\mathcal{G}}^\text{tie}_i (\tilde{\bm{b}}^\text{tie}_i )$ and $\hat{\mathcal{E}}_{t}^\text{bd} (\tilde{\bm{E}}_{t}^\text{bd}, \bm{e}_{t}^\text{bd} )$ that make up $\tilde{\mathcal{R}}_\text{L}$.
In each iteration, the parameters are updated to shrink the flexibility region. 
Denote the parameters in the $k$-th iteration as $\bm{b}^{\text{tie}}_{i(k)}$ and $\bm{E}_{t(k)}^{\text{bd}}$, and the flexibility region in the $k$-th iteration as $\hat{\mathcal{R}}_{\text{L}(k)}$. 
The iteration stops until the condition $\hat{\mathcal{R}}_{\text{L}(k)} \subseteq \mathcal{R}_\text{L}$ is met.
The initial values of the parameters in the iteration can be obtained in the previous dimension recombination step, denoted $\bm{b}^{\text{tie}}_{i(0)} := \tilde{\bm{b}}^\text{tie}_i$ and $\bm{E}_{t(0)}^{\text{bd}} := \tilde{\bm{E}}_{t}^\text{bd}$.

At the beginning of each iteration, the potential outliers, denoted as $\bm{x}^\text{out}_{(k)}$, such that $\bm{x}^\text{out}_{(k)} \in \hat{\mathcal{R}}_{\text{L}(k)}$ but $\bm{x}^\text{out}_{(k)} \notin \mathcal{R}_\text{L}$, are identified by solving a Stackelberg game \cite{chenLeveragingTwoStageAdaptive2021,wangAggregateFlexibilityVirtual2021} as follows:

\begin{subequations} \label{eq-MinmaxOpt}
	\begin{align}
		& \min_{\bm{x}^{\text{bd}}} \max_{\bm{x}^{\text{int}}} 0 \\
		\text{ s.t. } \quad & \bm{H}_{t}^\text{bd} \bm{x}^\text{bd} \in \hat{\mathcal{E}}_{t}^\text{bd}\left(\bm{E}_{t(k)}^{\text{bd}}, \bm{e}_{t}^\text{bd}\right), \forall t \in \mathcal{T} \\
		& \bm{A}^\text{tie}_i \bm{S}^\text{tie}_i \bm{x}^\text{bd} \leq \bm{b}^{\text{tie}}_{i(k)}, \forall i \in \mathcal{I}^\text{tie} \\
		& \bm{x}^\text{bd} = \bm{C} \bm{x}^\text{int} + \bm{d} \label{eq-SysEquality} \\
		& \bm{F} \bm{x}^\text{int} \leq \bm{f} \label{eq-SysInequality}
	\end{align}
\end{subequations}

\noindent In this Stackelberg game, the leader variables, $\bm{x}^{\text{bd}}$, aim to guarantee the feasibility of the problem, while the follower variables, $\bm{x}^{\text{int}}$, aim to make it infeasible. 
If the follower's maximization problem is infeasible, it indicates the existence of an outlier $\bm{x}^\text{out}_{(k)}$, such that $\bm{x}^\text{out}_{(k)} \in \hat{\mathcal{R}}_{\text{L}(k)}$ but $\bm{x}^\text{out}_{(k)} \notin \mathcal{R}_\text{L}$.
In this case, the optimal objective value of \eqref{eq-MinmaxOpt} is $-\infty$. 
Otherwise, if the follower's maximization problem is feasible, no such outlier exists, confirming that $\hat{\mathcal{R}}_{\text{L}(k)} \subseteq \mathcal{R}_\text{L}$, and the optimal objective value is 0. 
The problem \eqref{eq-MinmaxOpt} can be converted into an MILP formulation and solved, as described in detail in the supplementary file \cite{wangSupplementalFileNonIterative2024}.

If an outlier $\bm{x}^\text{out}_{(k)}$ is located by solving \eqref{eq-MinmaxOpt}, the nearest point on the boundary of $\mathcal{R}_\text{L}$, denoted as $\bm{x}^\text{bnd}_{(k)} \in \partial\mathcal{R}_\text{L}$, can be obtained by solving the following optimization problem:

\begin{subequations} \label{eq-nearestBound}
	\begin{align}
		\bm{x}^\text{bnd}_{(k)} & = \underset{\bm{x}^\text{bd}}{\operatorname{argmin}} \left\|\bm{x}^\text{out}_{(k)} - \bm{x}^\text{bd}\right\| \\
		\text{ s.t. } \quad & \left[ (\bm{x}^\text{bd})^\top (\bm{x}^\text{int})^\top \right]^\top \in \mathcal{R}_{\text{H}}
	\end{align}
\end{subequations}

After identifying the outlier $\bm{x}^\text{out}_{(k)}$ and the nearest boundary point $\bm{x}^\text{bnd}_{(k)}$, the parameters are modified to shrink the related bounds of $\hat{\mathcal{R}}_{\text{L}(k)}$ to draw back the outlier $\bm{x}^\text{out}_{(k)}$ to $\bm{x}^\text{bnd}_{(k)}$, as shown in Fig.~\ref{fig-boundShrinkAlgorithm}(c).
The index set of related bounds that $\bm{x}^\text{out}_{(k)} \in \partial\hat{\mathcal{R}}_{\text{L}(k)}$ can be identified as follows:

\begin{subequations} \label{eq-RelatedBounds}
	\begin{align}
		& \mathcal{T}_{(k)} := \left\{
			t \in \mathcal{T}
			\left\lvert
				\bm{H}^\text{bd}_t \bm{x}^\text{out}_{(k)} \in \partial\hat{\mathcal{E}}_{t}^\text{bd}(\bm{E}_{t(k)}^{\text{bd}}, \bm{e}_{t}^\text{bd})
			\right.
		\right\} \\
		& \mathcal{I}^\text{tie}_{(k)} := \left\{
			i \in \mathcal{I}^\text{tie}
			\left\lvert
				\bm{S}^\text{tie}_i \bm{x}^\text{out}_{(k)} \in \partial\hat{\mathcal{G}}^\text{tie}_i (\bm{b}^\text{tie}_{i(k)})
			\right.
		\right\} \\
		& \mathcal{J}_{i(k)} := \left\{
			j 
			\left\lvert
				(\bm{A}^\text{tie}_i)_j \bm{S}^\text{tie}_i \bm{x}^\text{out}_{(k)} = (\bm{b}^{\text{tie}}_{i(k)})_j
			\right.
		\right\}, \forall i \in \mathcal{I}^\text{tie}_{(k)}
	\end{align}
\end{subequations}

Subsequently, the bounds of $\hat{\mathcal{R}}_{\text{L}(k)}$ are shrunk by adjusting the parameters $\bm{E}_{t(k)}^\text{bd}$ and $\bm{b}^{\text{tie}}_{i(k)}$.

For regions inner-approximated with an ellipsoidal shape template, the radius is shrunk by scaling the shape matrix $\bm{E}_{t(k)}^\text{bd}$ with a factor $\rho_t$, while the center point vector $\bm{e}_{t}^\text{bd}$ remains fixed, $\forall t \in \mathcal{T}_{(k)}$:

\begin{subequations} \label{eq-EllipBoundShrink}
	\begin{align}
		& \bm{E}_{t(k+1)}^\text{bd} = \rho_{t} \bm{E}_{t(k)}^\text{bd}, \\
		& \rho_{t} := \left\|(\bm{E}_{t(k)}^\text{bd})^{-1}(\bm{H}^\text{bd}_t \bm{x}^\text{bnd}_{(k)}-\bm{e}_{t}^\text{bd})\right\|
	\end{align}
\end{subequations}

For regions inner-approximated with a polytopal shape template, the bounds are shrunk by adjusting the right-hand side parameters, $\bm{b}^{\text{tie}}_{i(k)}$, $\forall i \in \mathcal{I}^\text{tie}_{(k)}$:

\begin{equation} \label{eq-PolyBoundShrink}
	(\bm{b}^{\text{tie}}_{i(k+1)})_j = \left\{
		\begin{array}{l}
			(\bm{A}^\text{tie}_i)_j \bm{S}^\text{tie}_i \bm{x}^\text{bnd}_{(k)}, \text{if} \: j \in \mathcal{J}_{i(k)} \\
			(\bm{b}^{\text{tie}}_{i(k)})_j, \text{if} \: j \notin \mathcal{J}_{i(k)}
		\end{array}
	\right.
\end{equation}

Then update $k \leftarrow k+1$ and solve \eqref{eq-MinmaxOpt} again to check for any outliers in $\hat{\mathcal{R}}_{\text{L}(k+1)}$.

The pseudocode of the dimension-decomposition-based bound shrinking algorithm used to adjust the parameters is summarized in Algorithm \ref{alg-boundShrink}.

\begin{algorithm}[h]
	\caption{Dimension-Decomposition-Based Bound Shrinking Algorithm} \label{alg-boundShrink}
	\begin{algorithmic}[1]
		\STATE Initialize bound shrinking using the parameters obtained from dimension decomposition and set $k=0$.
		\STATE Solve \eqref{eq-MinmaxOpt} to check for the existence of an outlier $\bm{x}^\text{out}_{(k)} \in \hat{\mathcal{R}}_{\text{L}(k)}$ but $\bm{x}^\text{out}_{(k)} \notin \mathcal{R}_{\text{L}}$. If the optimal objective value of \eqref{eq-MinmaxOpt} is 0, exit the algorithm and output $\hat{\mathcal{R}}_{\text{L}(k)}$ as the final result; otherwise, proceed to Step 3.
		\STATE Solve \eqref{eq-nearestBound} to find the nearest point $\bm{x}^\text{bnd}_{(k)} \in \partial\mathcal{R}_\text{L}$.
		\STATE Solve \eqref{eq-RelatedBounds} identify which bounds of $\hat{\mathcal{R}}_{\text{L}(k)}$ the outlier $\bm{x}^\text{out}_{(k)}$ lies on.
		\STATE Shrink the related bounds of $\hat{\mathcal{R}}_{\text{L}(k)}$ by adjusting the parameters as in \eqref{eq-EllipBoundShrink} and \eqref{eq-PolyBoundShrink}. 
		\STATE Update $k \leftarrow k+1$ and return to Step 2.
	\end{algorithmic}
\end{algorithm}

The outputs of this algorithm are the adjusted vector $\bm{b}^\text{tie}_i$ and matrix $\bm{E}_{t}^\text{bd}$ of the inner-approximated flexibility region $\hat{\mathcal{R}}_\text{L}$.
Finally, the aggregated flexibility region of RPG is obtained as follows:

\begin{equation} \label{eq-RPGFlexRecombinedDetail}
	\hat{\mathcal{R}}_\text{L}:=\left\{
		\bm{x}^\text{bd} 
		\left\lvert\,
		\begin{array}{c}
			\bm{p}^\text{tie}_i := \bm{S}^\text{tie}_i \bm{x}^\text{bd} \in \hat{\mathcal{G}}^\text{tie}_i,
			\forall i \in \mathcal{I}^\text{tie} \\
			\bm{x}_{t}^\text{bd} := \bm{H}^\text{bd}_t \bm{x}^\text{bd} \in \hat{\mathcal{E}}_{t}^\text{bd},
			\forall t \in \mathcal{T}
		\end{array}
		\right.
	\right\}
\end{equation}

\noindent where

\begin{subequations}
	\begin{align}
	& \hat{\mathcal{G}}^\text{tie}_i := \left\{
		\bm{p}^\text{tie}_i 
		\left\lvert\,
			\bm{A}^\text{tie}_i \bm{p}^\text{tie}_i \leq \bm{b}^\text{tie}_i
		\right.
	\right\} \\
	& \hat{\mathcal{E}}_{t}^\text{bd} :=
	\left\{
		\bm{x}_{t}^\text{bd} 
		\left\lvert\,
			\left\|\left(\bm{E}_{t}^\text{bd}\right)^{-1}\left(\bm{x}_{t}^\text{bd}-\bm{e}_{t}^\text{bd}\right)\right\| \leq 1
		\right.
	\right\}
	\end{align}
\end{subequations}

The whole dimension-decomposition-based flexibility aggregation algorithm is summarized in Algorithm \ref{alg-dimDecomp}.

\begin{algorithm}[h]
	\caption{Dimension-Decomposition-Based Flexibility Aggregation for a Regional Power Grid} \label{alg-dimDecomp}
	\begin{algorithmic}[1]
		\STATE Analyze the temporal and spatial coupling relationships of boundary variables $\bm{x}^\text{bd}$. Categorize all the boundary variables into two types: 
		\begin{itemize}
			\item Temporal coupled variables $\bm{p}^\text{tie}_i, \forall i \in \mathcal{I}^\text{tie}$ 
			\item Single time slice coupled variables $\bm{x}^\text{bd}_t, \forall t \in \mathcal{T}$
		\end{itemize}
		
		\STATE Decompose the flexibility region into a series of lower-dimensional subspaces $\operatorname{space}(\bm{p}^\text{tie}_i), \forall i \in \mathcal{I}^\text{tie}$ and $\operatorname{space}(\bm{x}^\text{bd}_t), \forall t \in \mathcal{T}$. Apply shape templates to inner-approximate the projected regions in sub-spaces: 
		\begin{itemize}
			\item Use the polytopal template $\hat{\mathcal{G}}^\text{tie}_i$ \eqref{eq-RPGPolyMatrix} for $\operatorname{space}(\bm{p}^\text{tie}_i)$
			\item Use the ellipsoidal template $\hat{\mathcal{E}}_{t}^\text{bd}$ \eqref{eq-RPGEllipseInnerApprox} for $\operatorname{space}(\bm{x}^\text{bd}_t)$
		\end{itemize}
		Obtain the corresponding parameters $\tilde{\bm{b}}^\text{tie}_i, \forall i \in \mathcal{I}^\text{tie}$ and $\tilde{\bm{E}}_{t}^\text{bd}, \bm{e}_{t}^\text{bd}, \forall t \in \mathcal{T}$, respectively.
		
		\STATE Reconstruct the inner-approximated flexibility region in the full space $\operatorname{space}(\bm{x}^\text{bd})$ by recombining the sub-regions $\hat{\mathcal{G}}^\text{tie}_i, \forall i \in \mathcal{I}^\text{tie}$ and $\hat{\mathcal{E}}_{t}^\text{bd}, \forall t \in \mathcal{T}$ according to \eqref{eq-RPGFlexRecombined}.
		Obtain the corresponding parameters $\tilde{\bm{b}}^\text{tie}_i, \forall i \in \mathcal{I}^\text{tie}$ and $\tilde{\bm{E}}_{t}^\text{bd}, \bm{e}_{t}^\text{bd}, \forall t \in \mathcal{T}$.
		\STATE Adjust the parameters $\tilde{\bm{b}}^\text{tie}_i, \forall i \in \mathcal{I}^\text{tie}$ and $\tilde{\bm{E}}_{t}^\text{bd}, \forall t \in \mathcal{T}$ using Algorithm \ref{alg-boundShrink} to ensure $\hat{\mathcal{R}}_\text{L} \subseteq \mathcal{R}_\text{L}$ is satisfied. 
		Output the adjusted parameters as $\bm{b}^\text{tie}_i, \forall i \in \mathcal{I}^\text{tie}$ and $\bm{E}_{t}^\text{bd}, \forall t \in \mathcal{T}$.
		\STATE Return the final aggregated flexibility region for the regional power grid as defined in \eqref{eq-RPGFlexRecombinedDetail}.
	\end{algorithmic}
\end{algorithm}

\subsection{Physical Counterpart of the Flexibility Aggregation Model}

\begin{figure}[h] 
	\centering 
	\includegraphics[width=0.48\textwidth]{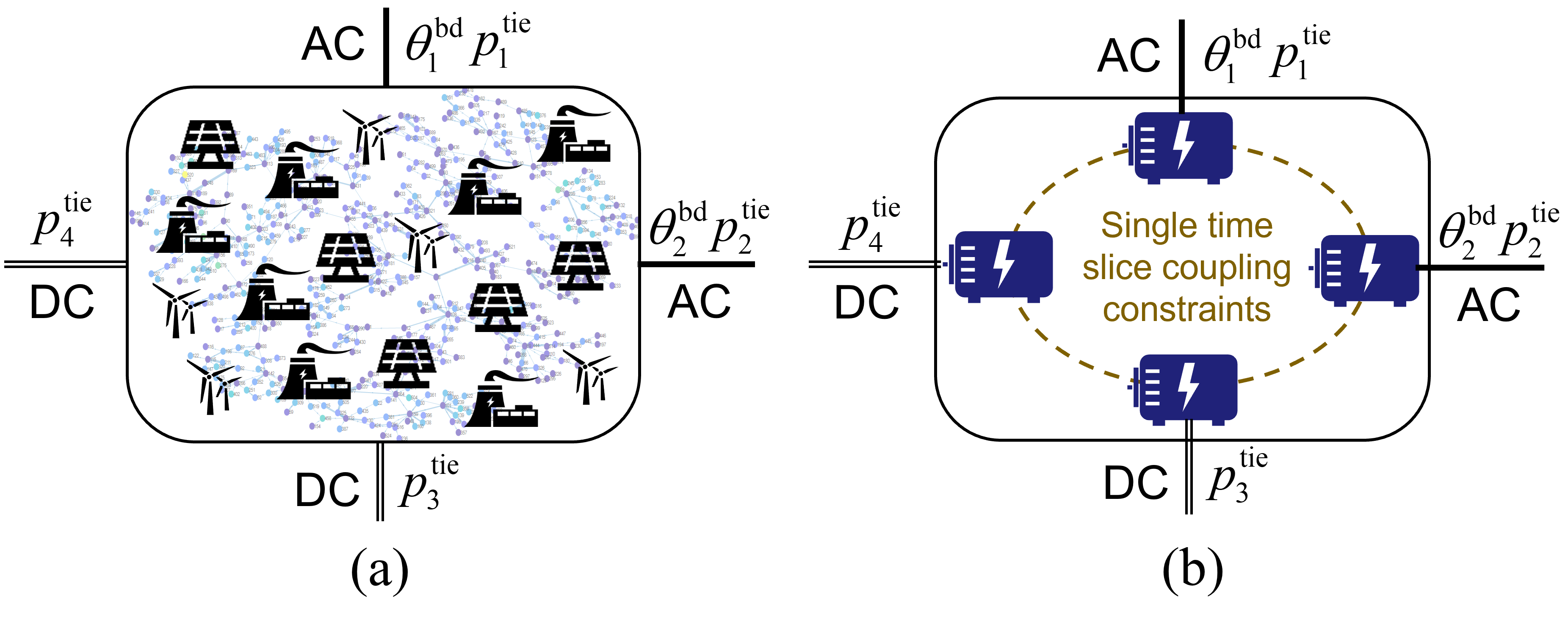}
	\caption{(a) Schematic diagram of the original regional power grid; (b) Physical counterpart: equivalent generators model via flexibility aggregation.} 
	\label{fig-aggRPGtoGens}
\end{figure}

The flexibility aggregation model derived through geometric projection calculations can be interpreted in terms of physical significance, as the parameters of the aggregated flexibility region of the RPG have a clear physical meaning, as shown in the schematic diagram Fig.~\ref{fig-aggRPGtoGens}.

For each single tie-line, its flexibility region across all time slots can be represented by an equivalent generator model, constrained by both power and ramp rate constraints, corresponding to the temporal coupling constraints in \eqref{eq-RPGPolyInnerApprox}.

For each single time slice, the output power and bus voltage angles of all equivalent generators are interconnected through the constraints in the ellipsoidal expression form, corresponding to the single time slice coupling constraints in \eqref{eq-RPGEllipseInnerApprox}.

Therefore, after flexibility aggregation, the entire RPG can be equivalent to a group of generators that interconnected through coupling constraints.

\begin{figure}[h] 
	\centering 
	\includegraphics[width=0.18\textwidth]{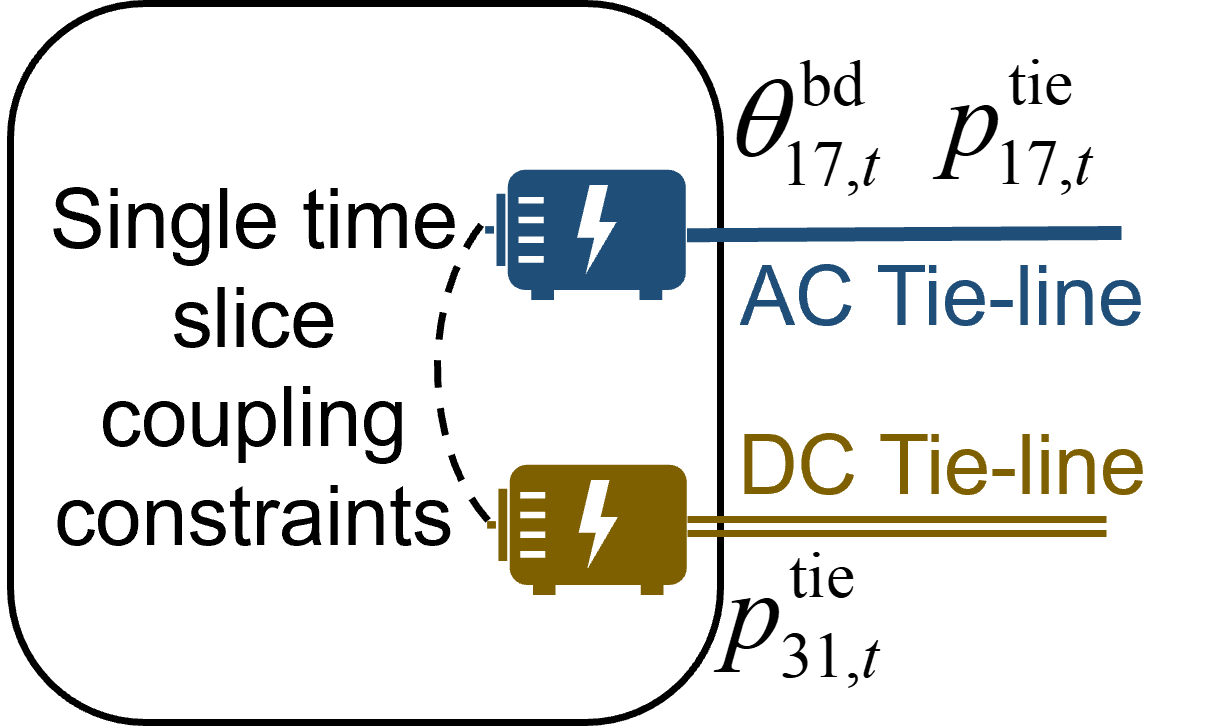}
	\caption{Physical counterpart of IEEE 118 regional power grid: two equivalent generators model and coupling constants.} 
	\label{fig-case118RPGtoGens}
\end{figure}

Taking the IEEE 118 RPG as an example, its physical counterpart of the flexibility aggregation model is composed of three parts: the equivalent generator of AC tie-line at Bus 17, the equivalent generator of DC tie-line at Bus 31, and the equivalent constraints among the boundary variables at each time slice, as shown in Fig.~\ref{fig-case118RPGtoGens}.
Each equivalent generator can be modeled as \eqref{eq-case118EquGen}. For $i \in \{17,31\}$

\begin{subequations} \label{eq-case118EquGen}
	\begin{align}
	& \underline{p}^\text{tie}_{i,t} \leq p^\text{tie}_{i,t} \leq \overline{p}^\text{tie}_{i,t}, \forall t \in \mathcal{T} \\
	& \underline{r}^\text{tie}_{i,t} \leq (p^\text{tie}_{i,t+1} - p^\text{tie}_{i,t}) / \Delta t \leq \overline{r}^\text{tie}_{i,t}, \forall t \in \mathcal{T}
	\end{align}
\end{subequations}

\noindent In addition, the equivalent constraints among the boundary variables at each time slice are in the ellipsoidal expression form as \eqref{eq-case118EquTimeSclice}.
For $t \in \mathcal{T}$

\begin{equation} \label{eq-case118EquTimeSclice}
	\left\|\left(\bm{E}_{t}^\text{bd}\right)^{-1}\left(\bm{x}^\text{bd}_t-\bm{e}_{t}^\text{bd}\right)\right\| \leq 1
\end{equation}

\noindent where $\bm{x}^\text{bd}_t := [p^\text{tie}_{17,t}, p^\text{tie}_{31,t}, \theta^\text{bd}_{17,t}]^\top$.

\subsection{Calculation of the Aggregated Operational Cost Function}

To facilitate coordinated economic dispatch of interconnected RPGs, the analytical aggregated cost function of RPG is derived with the piecewise linear fitting \cite{capitanescuComputingCostCurves2024}.
The total operational cost of an RPG is the sum of the operational costs of each generation unit. 
From an external system perspective, the operational cost of an RPG, denoted $\kappa_t$, is a function of the power flow through all tie-lines, denoted as $\bm{p}^\text{tie}_t$. 
To model this aggregated cost function, uniform sampling is applied in each dimension of the feasible domain, resulting in a total of $J$ samples of tie-line power flows and their corresponding operational costs.
The operating point and the corresponding cost of the $j$-th sample are denoted by $\bm{p}^\text{tie}_{t(j)}$ and $\kappa_{t(j)}$, respectively.
By solving the following optimal power flow problem, the minimum operational cost $\kappa_{t(j)}$ for the $j$-th operating point $\bm{p}^\text{tie}_{t(j)}$ can be obtained:

\begin{subequations} \label{eq-RPGAggCostFunc}
	\begin{align}
		\kappa_{t(j)}= & \underset{\bm{p}^\text{tpg}_t}{\operatorname{min}}\sum_{i \in \mathcal{I}^\text{tpg}} \left(\begin{array}{l}
				a^\text{tpg}_i (p^\text{tpg}_{i,t})^2 + b^\text{tpg}_i p^\text{tpg}_{i,t} + c^\text{tpg}_i \\
				+ d^\text{tpg}_i Ru^\text{tpg}_{i,t} + e^\text{tpg}_i Rd^\text{tpg}_{i,t}
			\end{array}\right) \\
		\text{ s.t. } \quad & \bm{p}^\text{tie}_{t} = \bm{p}^\text{tie}_{t(j)} \\
		& \text{Constraints} \, \eqref{eq-RPGCons}-\eqref{eq-RPGN1Cons} \notag
	\end{align}
\end{subequations}

\noindent where $a^\text{tpg}_i \sim e^\text{tpg}_i$ are the cost coefficients of the $i$-th thermal power generators in the RPG.

Subsequently, the convex piecewise linear fitting algorithm \cite{magnaniConvexPiecewiselinearFitting2009} is applied to fit these $J$ sample points into an analytical expression representing the aggregated operational cost function, denoted as

\begin{equation} \label{eq-RPGAggCostFuncPiecewise}
	\kappa_t\left(\bm{p}^\text{tie}_t\right) = \underset{i \in [m_t]}{\operatorname{max}} \left\{ \bm{a}^\top_{t,i} \bm{p}^\text{tie}_t + b_{t,i} \right\}
\end{equation}

\noindent where $m_t$ denotes the number of partitions in the piecewise-linear fitting. $[m_t]$ denotes the set of integers from 1 to $m_t$. 

Notably, the RPG may have numerous tie-lines, making it challenging to exhaustively cover all combinations of tie-line power flows and leading to the curse of dimensionality. 
In such cases, summing the power flows of all lines linking the same two RPGs and treating the combined power flow as one variable \cite{linHighdimensionTielineSecurity2023} in the cost function can effectively reduce the computational load for sampling and piecewise linear fitting in high-dimensional spaces.

\begin{figure}[h]
	\centering 
	\includegraphics[width=0.4\textwidth]{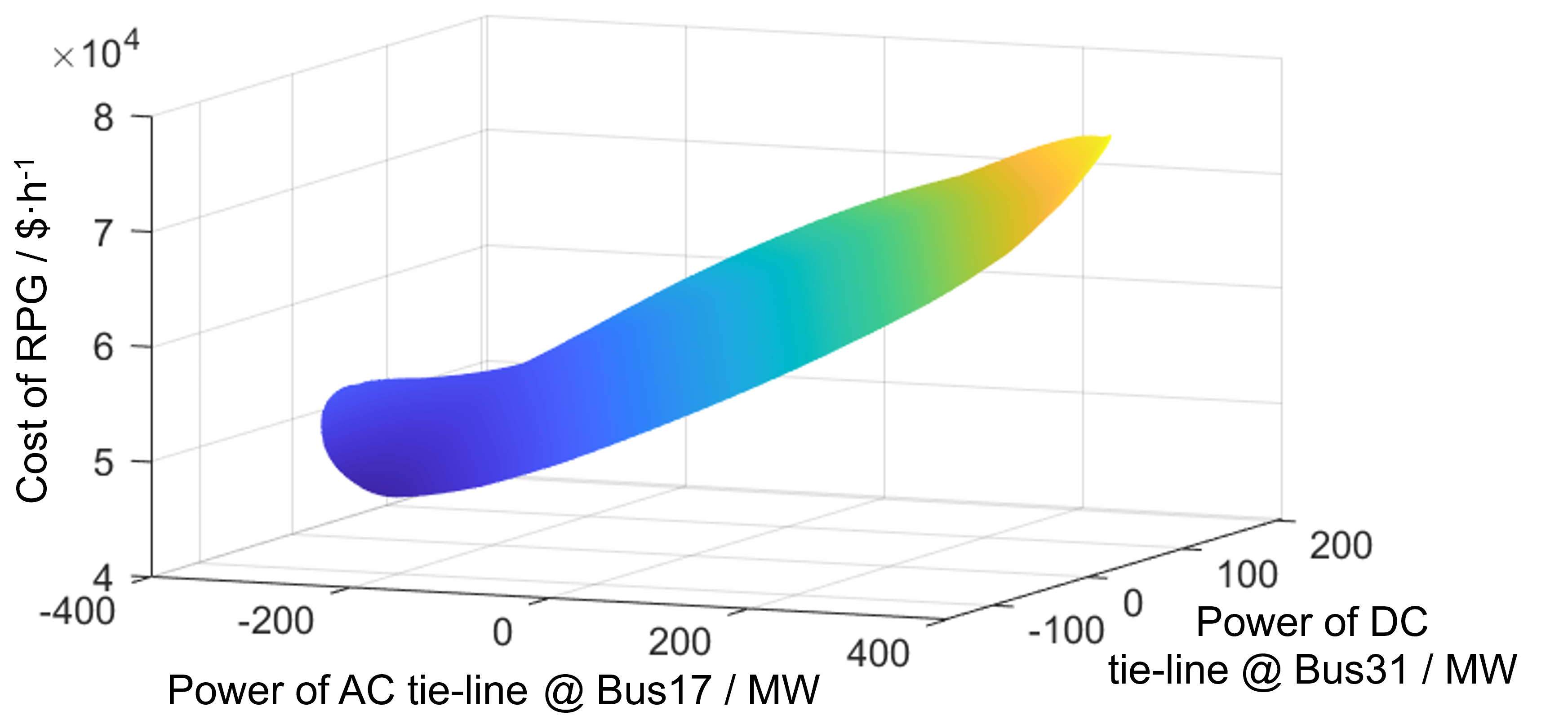}
	\caption{Aggregated cost function of IEEE 118 RPG.} 
	\label{fig-case118cost}
\end{figure}

In terms of the simple IEEE 118 case, the aggregated cost function with the output power of two tie-lines as variables at 12:00 is shown in Fig.~\ref{fig-case118cost}. 
The aggregated cost function is a piecewise linear function expressed by the power of two tie-lines with 64 partitions.

\section{Coordinated Dispatch of Interconnected RPGs Based on the Aggregated Model}
\label{sec-RPGAggCoordDispatch}

Using the aggregated flexibility models \eqref{eq-RPGFlexRecombinedDetail} and the aggregated cost function \eqref{eq-RPGAggCostFuncPiecewise} of each RPG, the coordinated dispatch across multiple interconnected RPGs can be formulated as

\begin{subequations} \label{eq-RPGAggCoordDispatch}
	\begin{align}
		\operatorname{min} & \sum_{r \in \mathcal{I}^\text{rg}} \sum_{t \in \mathcal{T}} \kappa_{r,t} \\
		\text{ s.t. } \quad & \forall t \in \mathcal{T}, \forall r \in \mathcal{I}^\text{rg}: \notag \\
		& \kappa_{r,t} \geq \bm{a}^\top_{r,t,i} \bm{p}^\text{tie}_t + b_{t,i}, \forall i \in [m_{r,t}] \label{eq-RPGAggCostFuncPiecewiseRelax} \\
		& \bm{x}^\text{bd}_r \in \hat{\mathcal{R}}_{\text{L}r}, -\overline{\bm{p}}^\text{tie}_{r} \leq \bm{p}_{r,t}^\text{tie} \leq \overline{\bm{p}}^\text{tie}_{r} \\
		& \text{Constraints} \: \eqref{eq-RPGTieLine} \notag
	\end{align}
\end{subequations}

\noindent where the subscript $r \in \mathcal{I}^\text{rg}$ represents the index of each RPG, and $\mathcal{I}^\text{rg}$ is the index set of RPGs.
$\overline{\bm{p}}^\text{tie}_{r}$ denotes the maximum capacity of the tie-lines connecting the $r$-th RPG to other RPGs. 
\eqref{eq-RPGAggCostFuncPiecewiseRelax} represents the convex relaxation of the piecewise linear cost function \eqref{eq-RPGAggCostFuncPiecewise}.
By solving \eqref{eq-RPGAggCoordDispatch}, the optimal scheduling of tie-lines of the interconnected RPGs can be obtained, which can be further used for the internal optimal dispatch of each RPG, respectively.


\section{Numerical Test}
\label{sec-NumericalTest}

\subsection{Simulation Setup}

We perform simulations on a five-region European grid to demonstrate coordinated-dispatch results using our dimension-decomposition-based flexibility aggregation methods.
This five-region power grid is spliced together by five European transmission systems sourced from the MATPOWER software package \cite{zimmermanMATPOWERSteadyStateOperations2011}. 
The five-region European power grid is shown in Fig.~\ref{fig-case5region}, with the detailed parameters listed in the supplementary file \cite{wangSupplementalFileNonIterative2024}.
The load curves for simulation are taken from California ISO historical data \cite{CaliforniaISOHistorical}, and the renewable energy generation curves are obtained from the NREL website \cite{jagerNRELNationalWind1996}.

\begin{figure}[h] 
	\centering 
	\includegraphics[width=0.4\textwidth]{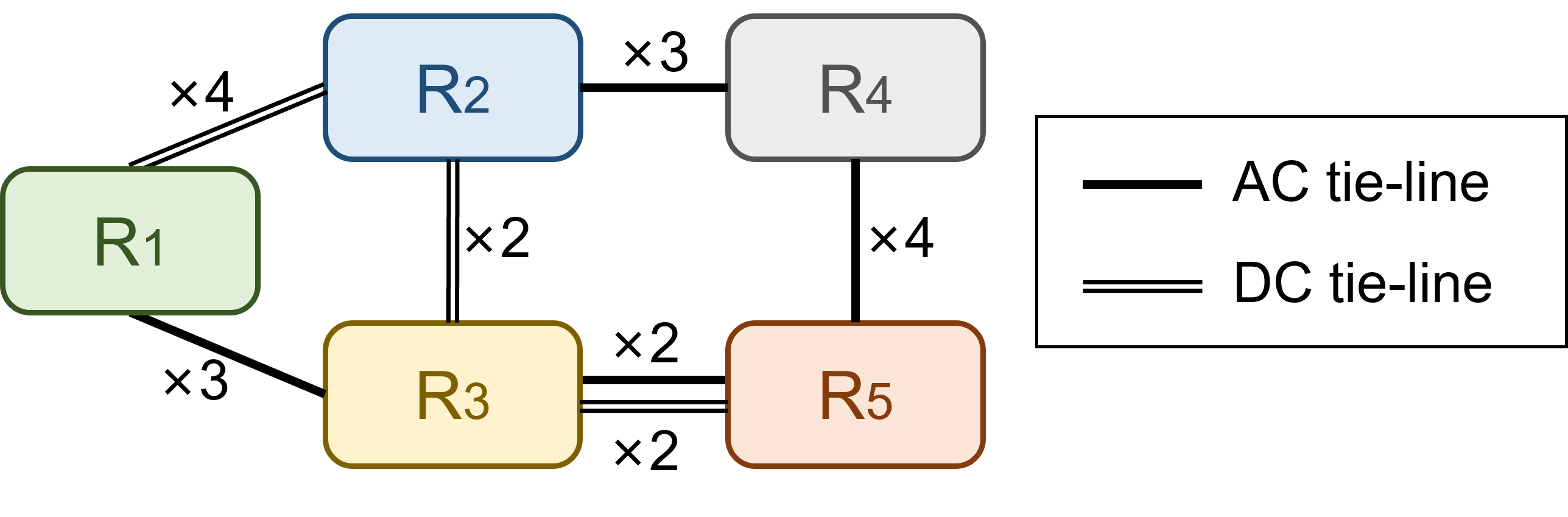}
	\caption{The five-region interconnected power grid.} 
	\label{fig-case5region}
\end{figure}
 
All test cases are simulated on a desktop computer with an Intel i7-10700 CPU at 2.90GHz and 64 GB RAM, running on MATLAB software. 
The YALMIP optimization toolbox \cite{lofbergYALMIPToolboxModeling2004}, MPT3 \cite{hercegMultiParametricToolbox302013}, solvers COPT \cite{copt} and Gurobi \cite{gurobioptimizationllcGurobiOptimizerReference2024} are used to solve the optimization problems.

\subsection{Optimal Coordination Based on the Aggregated Models of RPGs}
\begin{figure}[h] 
	\centering 
	\includegraphics[width=0.48\textwidth]{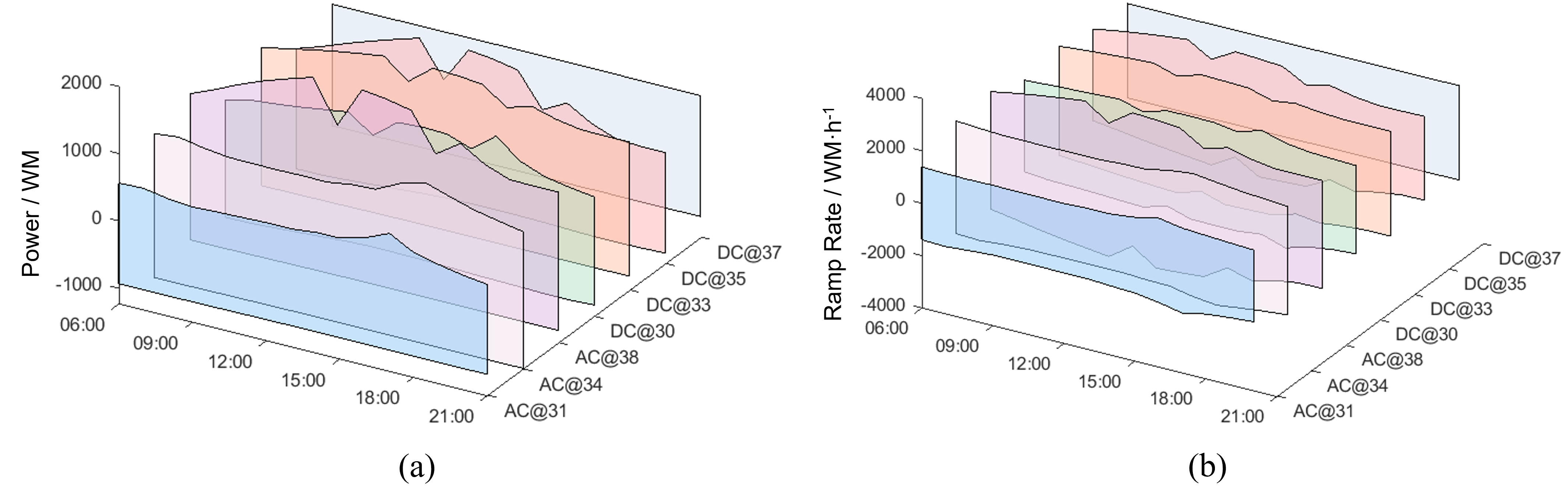}
	\caption{Aggregation results of RPG R1: (a) Power bounds; (b) Ramp rate bounds.} 
	\label{fig-aggR1Gens}
\end{figure}

Based on the proposed methodologies, the aggregated models of these five-region European RPGs can be calculated, respectively. 
Taking the RPG R1 as an example, the power and ramp rate bounds of its seven tie-lines are shown in Fig.~\ref{fig-aggR1Gens}.
These parameters can also be viewed as the parameters of the seven equivalent generators connected to these tie-lines like Fig.~\ref{fig-aggRPGtoGens}(b).

\begin{figure}[h] 
	\centering 
	\includegraphics[width=0.43\textwidth]{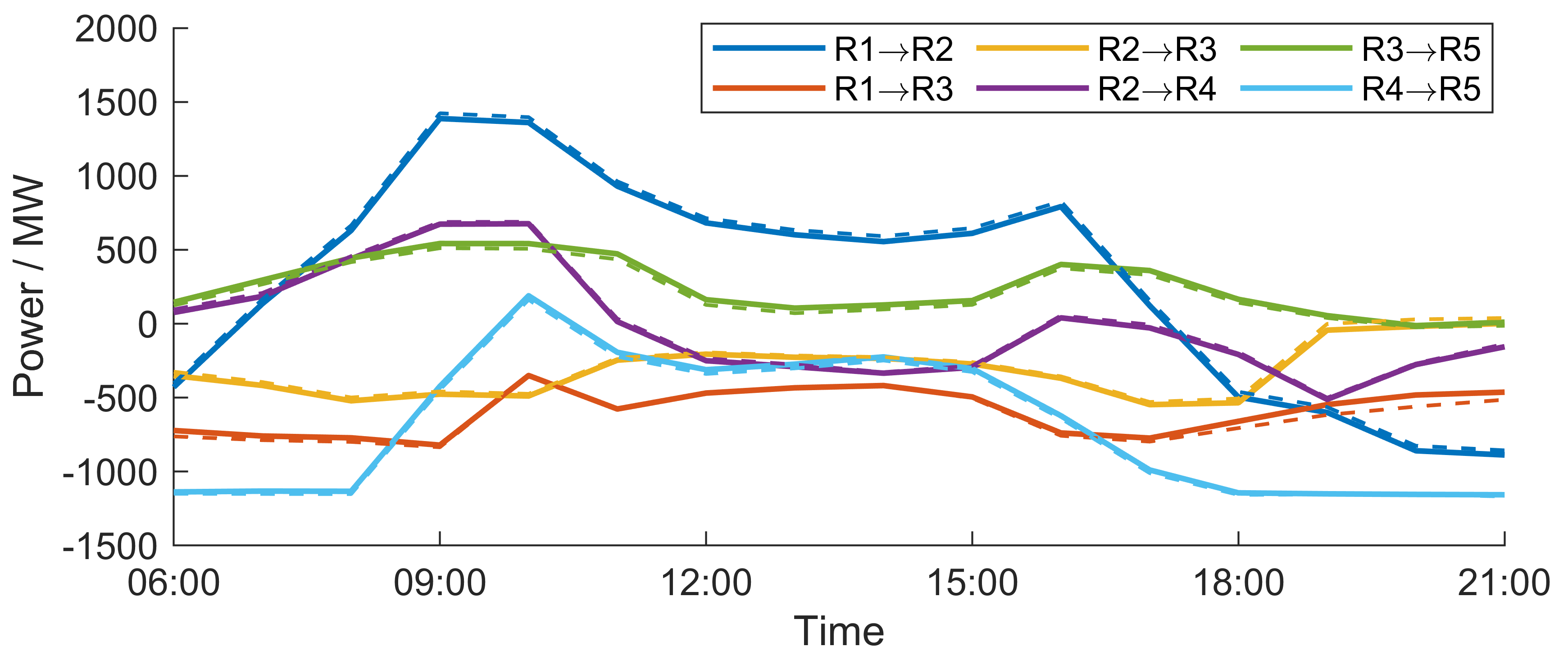}
	\caption{Coordinated dispatch results of the five-region European power grid.} 
	\label{fig-multiRegionDispatch}
\end{figure}

Using the flexibility aggregation models of these five RPGs, the coordinated dispatch results for the interconnected power grid can be achieved in a non-iterative manner. 
Fig.~\ref{fig-multiRegionDispatch} illustrates the power transmission between RPGs via tie-lines. 
The dotted lines represent the coordinated results obtained using the flexibility aggregation models in non-iterative method, while the solid lines depict the results from the distributed iterative dispatch.

According to Fig.~\ref{fig-multiRegionDispatch}, errors are present when using the flexibility aggregation models in non-iterative approach.
The errors mainly come from two aspects: the inner approximation errors between the flexibility aggregation models and the actual flexibility region, and the mismatch between the fitted cost function and the actual cost function.
Despite these inaccuracies, the flexibility aggregation model-based coordination results can provide a good initial value for further iterative method's distributed coordination calculations. 
This accelerates the convergence of the distributed algorithm and helps mitigate the risk of divergence in the distributed coordinated scheduling process.

\subsection{Comparison with Existing Aggregation Methods}

In this section, the performance of the proposed method is compared with the existing methods, including Fourier-Motzkin elimination \cite{a.a.jahromiLoadabilitySetsPower2017}, vertex search \cite{tanEnforcingIntraRegionalConstraints2019}, multi-parametric programming \cite{daiSecurityRegionRenewable2019}, cubic inner approximation \cite{chenLeveragingTwoStageAdaptive2021} and temporal dimension decomposition \cite{linHighdimensionTielineSecurity2023}.

In terms of algorithmic complexity, let $T$ denote the total number of time slots considered. The Fourier-Motzkin elimination method exhibits a computational complexity of $O(T^{2^T})$, while both the vertex enumeration approach and multi-parametric programming have complexity on the order of $O(T^T)$. The others exhibit exponential complexity with respect to $O(2^T)$.

To compare the performance of the aggregated flexibility region of the RPG, we adopt the scenario coverage rate as an evaluation metric. 
This metric is estimated using a Monte Carlo method, which approximates the volume ratio between the inner-approximated flexibility region $\hat{\mathcal{R}}_\text{L}$ and the true feasible region $\mathcal{R}_\text{L}$. 
In this test case, it is defined as the proportion of 100,000 randomly generated operational scenarios that satisfy both technical constraints and are included within the aggregated flexibility region. 
The scenario coverage rate serves as an indicator of how closely the computed flexibility region approximates the actual feasible region.

Due to the high-complexity nature of this aggregation problem, methods like Fourier-Motzkin elimination, vertex search, and multi-parametric programming, suffer from the curse of dimensionality and are unable to complete the calculation within one hour.
The comparison results of other methods are shown in TABLE \ref{tab-comparision}.

\begin{table}[h]
	\centering
	\footnotesize
	\caption{Comparison of the Flexibility Aggregation Results}
	\renewcommand{\arraystretch}{1.5}
	\begin{tabularx}{0.45\textwidth}{>{\centering\arraybackslash}m{0.16\textwidth}>{\centering\arraybackslash}m{0.1\textwidth}>{\centering\arraybackslash}m{0.1\textwidth}}
        \hline\hline
        Method & Computation time / s & Scenario coverage rate \\ 
        \hline
        Cubic inner approximation \cite{chenLeveragingTwoStageAdaptive2021}  & 243 & 36.72\%  \\
        Temporal dimension decomposition \cite{linHighdimensionTielineSecurity2023} & 673 & 69.87\%  \\
        Our method  & 617 & 85.49\%  \\
        \hline\hline
	\end{tabularx} 
	\label{tab-comparision}
\end{table}

The cubic inner approximation method \cite{chenLeveragingTwoStageAdaptive2021} provides much faster computation speed by using a simple cubic shape template to inner-approximate the real flexibility region.
The temporal dimension decomposition method \cite{linHighdimensionTielineSecurity2023} and our proposed method, while slower, offer a more accurate representation of the aggregated flexibility region. 
In addition, compared to the time-dimension decomposition method, our proposed method in this paper achieves a more precise inner approximation of the flexibility region by incorporating the ramping characteristics of tie-lines.

\subsection{Scalability Analysis}

\begin{figure}[h] 
	\centering 
	\includegraphics[width=0.43\textwidth]{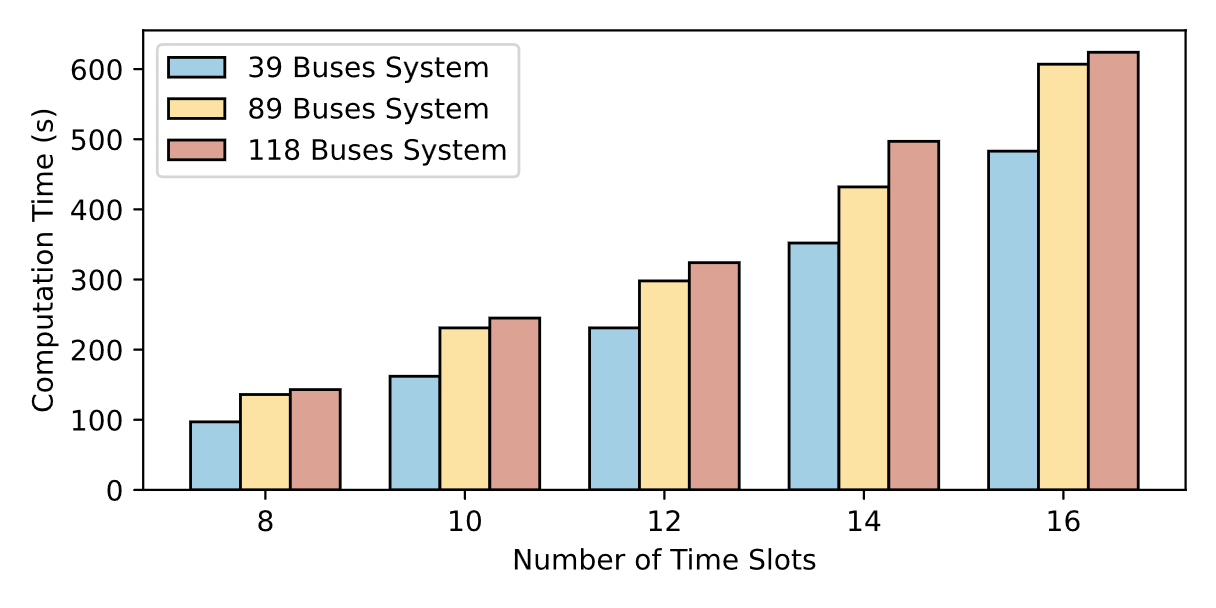}
	\caption{Computation time across varying grid sizes and different time slots.} 
	\label{fig-scalability}
\end{figure}

To evaluate the scalability of the proposed algorithm, we conducted experiments using three RPG instances with varying grid sizes and differing numbers of time slots. 
As shown in Fig.~\ref{fig-scalability}, computation time is more sensitive to the number of time slots than to the size of the grid itself. 
This is attributed to the fact that the number of boundary variables increases roughly in proportion to the number of time slots. 
These results corroborate our prior analysis, which suggests that the computational complexity increases nearly exponentially with the number of time slots.

\section{Conclusion}
\label{sec-Conclusion}

We propose a dimension-decomposition-based inner-approximation method to aggregate the flexibility of regional grids with multiple AC/DC tie-lines.
The aggregated model integrates the detail models of internal devices and operation constraints, facilitating optimal coordination across interconnected power grids in a non-iterative way.

To accelerate the calculation of the aggregated model, enhance coverage of the flexibility region, and derive physically meaningful results, we employ three key strategies:
(i) decomposing the high-dimensional flexibility aggregation region into subspaces based on the analysis of the coupling relationships among boundary variables; 
(ii) solving the projection problem within each subspace independently and in parallel; and
(iii) employing appropriate shape templates to inner-approximate the flexibility regions.

Numerical tests demonstrate that this flexibility aggregation method is significantly less conservative than existing methods while reliably ensuring the feasibility of the aggregated results in the dispatch process of each regional power grid.

\bibliographystyle{IEEEtran}
\bibliography{references}

\end{sloppypar}
\end{document}